\documentclass[aps,prl,amsmath,amssymb,amsfonts,nofootinbib,long,floatfix]{revtex4}
\usepackage{epsfig,latexsym,bm,longtable}

\newcommand\Mpc{\mbox{Mpc}}
\newcommand\keV{\mbox{keV}}
\newcommand\km{\mbox{km}}
\newcommand\s{\mbox{s}}

\begin{document}

\title{Galaxy cluster number count data constraints on cosmological parameters}


\author{L.\ Campanelli$^{1}$}
\email{leonardo.campanelli@ba.infn.it}

\author{G.\ L.\ Fogli$^{1,2}$}
\email{gianluigi.fogli@ba.infn.it}

\author{T.\ Kahniashvili$^{3,4,5}$}
\email{tinatin@phys.ksu.edu}

\author{A.\ Marrone$^{1,2}$}
\email{antonio.marrone@ba.infn.it}

\author{Bharat\ Ratra$^{6}$}
\email{ratra@phys.ksu.edu}

\affiliation{$^1$Dipartimento di Fisica, Universit\`{a} di Bari, I-70126 Bari, Italy}
\affiliation{$^2$INFN - Sezione di Bari, I-70126 Bari, Italy}
\affiliation{$^3$McWilliams Center for Cosmology and Department of Physics,
                 Carnegie Mellon University, 5000 Forbes Ave, Pittsburgh, PA 15213, USA}
\affiliation{$^4$Department of Physics, Laurentian University, Ramsey Lake Road, Sudbury, ON P3E 2C, Canada}
\affiliation{$^5$Abastumani Astrophysical Observatory, Ilia State University, 2A Kazbegi Ave, Tbilisi, GE-0160, Georgia}
\affiliation{$^6$Department of Physics, Kansas State University, 116 Cardwell Hall, Manhattan, KS 66506, USA}

\date{August 2012 \ \ \ KSUPT-11/4}


\begin{abstract}
\begin{center}
{\bf Abstract}
\end{center}
We use data on massive galaxy clusters ($M_{\rm cluster} > 8 \times 10^{14} h^{-1} M_\odot$
within a comoving radius of $R_{\rm cluster} = 1.5 h^{-1}\Mpc$) in the redshift range
$0.05 \lesssim z \lesssim 0.83$ to place constraints, simultaneously, on the nonrelativistic
matter density parameter $\Omega_m$, on the amplitude of mass fluctuations $\sigma_8$,
on the index $n$ of the power-law spectrum of the density perturbations, and on the Hubble
constant $H_0$, as well as on the equation-of-state parameters $(w_0,w_a)$ of a
smooth dark energy component.

For the first time, we properly take into account the dependence on redshift
and cosmology of the quantities related to cluster physics: the critical density contrast,
the growth factor, the mass conversion factor, the virial overdensity, the virial radius
and, most importantly, the cluster number count derived from the observational temperature data.

We show that, contrary to previous analyses, cluster data alone prefer low values of the amplitude of mass
fluctuations, $\sigma_8 \leq 0.69 \; (1\sigma \ \mbox{C.L.})$, and large amounts of
nonrelativistic matter, $\Omega_m \geq 0.38 \; (1\sigma \ \mbox{C.L.})$, in slight tension with
the $\Lambda$CDM concordance cosmological model, though the results are
compatible with $\Lambda$CDM at $2\sigma$. In addition, we derive a
$\sigma_8$ normalization relation,
$\sigma_8 \Omega_m^{\,1/3} = 0.49 \pm 0.06 \; (2\sigma \ \mbox{C.L.})$.

Combining cluster data with $\sigma_8$-independent baryon acoustic oscillation
observations, cosmic microwave background data, Hubble constant measurements,
Hubble parameter determination from passively-evolving red galaxies,
and magnitude-redshift data of type Ia supernovae,
we find $\Omega_m = 0.28^{+0.03}_{-0.02}$ and $\sigma_8 = 0.73^{+0.03}_{-0.03}$,
the former in agreement and the latter being slightly lower than the corresponding values
in the concordance cosmological model.
We also find $H_0 = 69.1^{+1.3}_{-1.5} \; \km/\s/\Mpc$,
the fit to the data being almost independent on $n$ in the adopted range $[0.90,1.05]$.

Concerning the dark energy equation-of-state parameters, we show that the present
data on massive clusters weakly constrain $(w_0,w_a)$ around the values corresponding to a cosmological constant,
i.e. $(w_0,w_a) = (-1,0)$. The global analysis gives
$w_0 = -1.14^{+0.14}_{-0.16}$ and $w_a = 0.85^{+0.42}_{-0.60}$
($1\sigma$ C.L.\ errors).
Very similar results are
found in the case of time-evolving dark energy with a constant equation-of-state parameter
$w = \mbox{const}$ (the XCDM parametrization). Finally, we show that the impact of bounds
on ($w_0,w_a$) is to favor top-down phantom models of evolving dark energy.
\end{abstract}


\maketitle



\section{\normalsize{I. Introduction}}
\renewcommand{\thesection}{\arabic{section}}

In the last few years galaxy cluster observations have begun to provide useful constraints
on cosmological parameters (for a recent review see Ref.~\cite{Review}). Data on
galaxy clusters are now used to test the validity of the standard cosmological model, the
so-called $\Lambda$ cold dark matter ($\Lambda$CDM) concordance model, \cite{Peebles84}, which
describes observational data at large cosmological scales (from galactic scales to the present
horizon scale) reasonably well \cite{Weinberg}. In particular, cluster observations can help
tighten the bounds on cosmological parameters such as the nonrelativistic matter density parameter
$\Omega_m$, the amplitude of mass fluctuations $\sigma_8$, the power-law index $n$ of the density
perturbation power spectrum, and the Hubble constant $H_0$~\cite{Voit}.

When combined with other cosmological probes --- such as cosmic microwave background (CMB)
radiation anisotropy, baryon acoustic oscillations (BAO) in the matter power spectrum,
Hubble parameter, and type Ia supernovae (SNeIa) data --- galaxy cluster observations provide
a unique insight towards helping understand the evolution of the Universe, from the inflation
era to today.

Despite the observational success of the $\Lambda$CDM model, a number of basic questions
remain unanswered. Dark energy is a major mystery (for reviews on dark energy and modified
gravity see, e.g., Refs.\ \cite{Dark}). A possibility is that dark energy is simply a
manifestation of a nonzero vacuum energy, a cosmological constant $\Lambda$,\footnote{
It has been known for some time that a spatially-flat $\Lambda$CDM model  is consistent
with most observational constraints, see, e.g., Refs.\ \cite{LambdaObs}. In the $\Lambda$CDM
model the energy budget is dominated by far by a cosmological constant,
a spatially homogenous fluid with equation of state parameter $w_\Lambda =
p_\Lambda/\rho_\Lambda = -1$ (where $p_\Lambda$ and $\rho_\Lambda$ are the fluid pressure
and energy density), with nonrelativistic CDM being the second largest
contributor. Note that the ``standard'' CDM structure formation model --- which the
standard $\Lambda$CDM cosmological model assumes --- might have some observational
inconsistencies (see, e.g., \cite{CDMtrouble}).}
but dynamical scalar field models of dark energy, \cite{PR88, RP88}, are also compatible
with present data.\footnote{
In dynamical dark energy models the dark energy density decreases in time and so
remains comparable to the nonrelativistic matter (CDM and baryons) energy density
for a longer time (than does a time-independent $\Lambda$). This partially
alleviates the ``coincidence'' puzzle of the $\Lambda$CDM model. In addition, some
dynamical dark energy scalar field models have a nonlinear attractor solution that
generates the current, tiny, dark energy density energy scale of order an meV from
a significantly higher energy density scale (possibly of a more fundamental model)
as a consequence of the very slow decrease in time of the dark energy density during
the very long age of the Universe. These results are often viewed as providing
significant theoretical motivation to consider dynamical dark energy models, \cite{PR88, RP88}.}

Measurements of the local abundance and growth of galaxy clusters from
X-ray \cite{Mantz1,Mantz2,Vikhlinin,K12} and optical \cite{Rozo} surveys have been
recently used to probe the standard cosmological model. In particular, the emerging
picture is that a cosmological constant still remains a good candidate for dark energy.
This conclusion does not exclude the possibility that future cluster surveys will allow
us to discriminate between the $\Lambda$CDM model and dynamical dark energy
models~\cite{Basilakos}.

The aim of this paper is to use present data on massive galaxy clusters
($M_{\rm cluster} > 8 \times 10^{14} h^{-1} M_\odot$ within a comoving radius of
$R_{\rm cluster} = 1.5 h^{-1}\Mpc$) at low and high redshifts ($0.05 \lesssim z
\lesssim 0.83$) to constrain some of the free parameters of the standard cosmological
model, and to investigate the possibility that the dark energy density evolves in time,
instead of staying constant.

It is worth noticing that the evolution with redshift of massive clusters is very sensitive
to the cosmological parameters, especially to $\sigma_8$ and $\Omega_m$~\cite{Bahcall}.
In particular, the abundance of massive clusters depends
exponentially on $\sigma_8$, in such a way that high values of $\sigma_8$
favor the formation of structures at early times, while a low amplitude of
mass fluctuations results in few massive clusters forming at high redshifts.

In Refs.\ \cite{Bahcall} these data were used to determine
the linear amplitude of mass fluctuations and the nonrelativistic matter density in a
Universe with a cosmological constant. We extend the analysis of Refs.\ \cite{Bahcall} to
the case of evolving dark energy, and we properly take into account the dependence on redshift
and cosmology of quantities related to cluster physics, such as the critical density contrast,
the growth factor, the mass conversion factor, the virial overdensity, and the virial radius.
Most importantly, we consider the dependence on redshift
and cosmology of the cluster number count derived from the observational data.
We emphasize that the observed number of clusters with masses
exceeding a fixed threshold is calculated as the number of clusters
with X-ray temperature larger than a corresponding temperature threshold,
and that the relation between mass and temperature depends on the redshift and
cosmological parameters.

It is of great interest to determine if the dark energy is well-approximated by
a cosmological constant
or if it decreases slowly in time (and so varies weakly in space). Ideally one would
very much prefer a model-independent resolution of this issue. However, at this point
in time, observational data are not up to this task. One must instead use the available
observational data to constrain model parameters and so determine if the cosmological
constant point in model parameter space is or is not favored over points where the
dark energy density slowly decreases in time.
While it is useful to perform such an analysis using a consistent and physically
motivated model, such as the inverse power-law potential energy density scalar
field model \cite{PR88, RP88}, this is computationally quite demanding, so here we
make use of a simple parametrization of time-evolving dark energy in a preliminary
attempt to investigate this matter.

In order to discriminate between a cosmological constant and dynamical
dark energy we use the dark energy equation-of-state parameter
parametrization \cite{CP}
\begin{equation}
\label{wz}
w(a) = w_0 + w_a \, (1-a) \ ,
\end{equation}
where $a$ is the scale factor related to the redshift $z$ by $a=1/(1+z)$.
The cosmological constant corresponds to $w_0 = -1$ and $w_a = 0$, the
case of constant equation of state corresponds to $w_0 = w = \mbox{const}$ and $w_a = 0$
(known as the XCDM parametrization of time-evolving dark energy), while the general case
of time-evolving dark energy corresponds to $w_a \neq 0$.

In this paper, we consider the case of a smooth dark energy component, namely 
the case where the dark energy does not cluster. For clustering dark energy, 
it could be expected that the bounds on dark energy
equation-of-state parameters $w_0,w_a$, as well as on the cosmological parameters $(\Omega_m, \sigma_8)$,
will be weakened because of the degeneracy between the above parameters and the effective dark energy sound speed
(which parameterizes the level of dark energy clustering) \cite{Abramo}.

The plan of our paper is as follows. In the next section, we introduce the basic theory and data on
galaxy cluster number counts used in our analysis. In Section III,
we present data from
other types of cosmological probes:
baryon acoustic oscillations, cosmic microwave background radiation anisotropies,
passively-evolving red galaxies, and type Ia supernovae.
In Section IV, we outline a joint analysis of all data
and discuss the results on the $(\Omega_m,\sigma_8)$ and $(w_0,w_a)$ planes, and
in the XCDM case.
In Section V,
we briefly discuss the impact of our results on some models
of evolving dark energy.
In Section VI, we draw our conclusions.
Finally, in the Appendices we discuss in detail
the critical density contrast and the growth factor (Appendix A),
the mass conversion factor (Appendix B), and the
virial overdensity and the virial radius (Appendix C).


\section{\normalsize{II. Galaxy Cluster Number Counts}}
\renewcommand{\thesection}{\arabic{section}}

In this section we introduce the basic physical quantities and observables related
to galaxy cluster number counts and we discuss the available experimental data.

\subsection{\normalsize{IIa. Theory}}
\renewcommand{\thesection}{\arabic{section}}

{\it Cluster number and comoving volume.}-- The comoving number of clusters in the
redshift interval $[z_1,z_2]$, whose mass $M$ is greater than a fiducial mass $M_0$, is
\begin{equation}
\label{Number}
\mathcal{N} = \int_{z_1}^{z_2} \! dz \, \frac{dV(z)}{dz} \, N(M>M_0,z) \ ,
\end{equation}
where
\begin{equation}
\label{V}
V(z) = 4 \pi \int_0^z \! dz' \, \frac{d_L^{\,2}(z')}{(1+z')^2H(z')}
\end{equation}
is the comoving volume at redshift $z$, and
\begin{equation}
\label{dL}
d_L(z) = (1+z) \int_0^z \frac{dz'}{H(z')}
\end{equation}
is the luminosity distance with $H(z)$ the Hubble parameter.
The ``mass function'' $N(M>M_0,z)$ appearing in Eq.~(\ref{Number})
represents the comoving cluster number density at redshift $z$ of clusters
with masses greater than $M_0$.

For a cosmological model with evolving dark energy equation-of-state parameter of the form~(\ref{wz}),
the Hubble parameter normalized to its present value $H_0$ is
\begin{equation}
\label{H}
E(z) = \frac{H(z)}{H_0} =
\left[ \frac{\rho_m(z)}{\rho_{\rm cr}^{(0)}} + \frac{\rho_{\rm DE}(z)}{\rho_{\rm cr}^{(0)}} \right]^{1/2} \ .
\end{equation}
The quantities
\begin{equation}
\label{rhom}
\rho_m(z) = \Omega_m \, \rho_{\rm cr}^{(0)} (1+z)^3
\end{equation}
and
\begin{equation}
\label{rhoDE}
\rho_{\rm DE}(z) = \Omega_{\rm DE} \, \rho_{\rm cr}^{(0)} (1+z)^{3(1+w_0+w_a)} \, e^{-3w_a z/(1+z)}
\end{equation}
are the energy densities of nonrelativistic matter and dark energy, respectively.
Here, $\Omega_m = \rho_m^{(0)}/\rho_{\rm cr}^{(0)}$ and
$\Omega_{\rm DE} = \rho_{\rm DE}^{(0)}/\rho_{\rm cr}^{(0)}$
are the matter and dark energy density parameters,
and $\rho_m^{(0)}$, $\rho_{\rm DE}^{(0)}$, and $\rho_{\rm cr}^{(0)} = 3H_0^2/(8\pi G)$
are the present matter, dark energy, and critical energy densities, respectively,
while $G$ is the Newton constant.

In this paper, for computational simplicity, we restrict ourselves to the case of a
flat universe,\footnote{
This is consistent with the simplest interpretation of the CMB anisotropy data, see, e.g.,
\cite{Podariu01, WMAP7}.}
so
\begin{equation}
\label{flat}
\Omega_m + \Omega_{\rm DE} = 1 \ .
\end{equation}

{\it Mass function.}-- To compute the mass function, we use the
Press-Schechter (PS) approach~\cite{PS}, as modified by Sheth and
Tormen (ST)~\cite{ST}. In this approach, the mass function is
written as
\begin{equation}
\label{N}
N(M>M_0,z) = \int_{M_0}^\infty \! dM \, n(M,z) \ ,
\end{equation}
and
\begin{equation}
M = \frac{4\pi}{3} \, r^3(z) \, \rho_m(z) = \frac{4\pi}{3} \, R^3 \rho_m^{(0)}
\end{equation}
is the mass within a sphere of physical radius $r(z)$, whose corresponding comoving
radius is $R = (1+z) \, r(z)$.

In Eq.~(\ref{N}), $n(M,z) \, dM$ is the comoving number
density at redshift $z$ of clusters with masses in the interval $[M,M+dM]$, and
is written as
\begin{equation}
\label{n}
n = \frac{2\rho_m}{M} \, \nu f(\nu) \, \frac{d\nu}{dM} \ .
\end{equation}
Here the multiplicity function $\nu f(\nu)$ is (in the PS and ST models)
an universal function of the peak height
\begin{equation}
\label{nu}
\nu = \frac{\delta_c}{\sigma} \ ,
\end{equation}
and is normalized as
\begin{equation}
\label{intf}
\int_0^{\infty} \! d\nu \, \nu f(\nu) = \frac12 \ .
\end{equation}
The functional form of $\nu f(\nu)$ is discussed below.
The critical density contrast $\delta_c(z)$ is the density
contrast for a linear overdensity able to collapse at the redshift $z$, and its
dependence on cosmological parameters is discussed in Appendix A.

The root mean square (rms) amplitude $\sigma$ of density fluctuations in a sphere
of comoving radius $R$, whose corresponding physical radius $r$ contains the mass $M$,
is related to the matter power spectrum of density perturbations at redshift $z$,
$P(k,z)$, through
\begin{equation}
\label{sigma}
\sigma^2(R,z) = \frac{1}{2\pi^2} \int_0^{\infty} \!\! dk k^2 P(k,z) W^2(kR) \ .
\end{equation}
Here
\begin{equation}
\label{W}
W(x) = \frac{3(\sin x - x \cos x)}{x^3}
\end{equation}
is the Fourier transform of the top-hat window function and
\begin{equation}
\label{P}
P(k,z) = P_0(k) \, T^2(k) D^2(z) \ ,
\end{equation}
with $D(z)$ being the growth factor (discussed in Appendix A) and
$T(k)$ the transfer function.

We assume that the post-inflationary density perturbation power
spectrum $P_0(k)$ is a simple power law,
\begin{equation}
\label{P0}
P_0(k) = A k^n \ ,
\end{equation}
with the scale-invariant spectrum corresponding to $n = 1$. The normalization
constant $A$ is a free parameter of the model and can be expressed as a
function of the other cosmological parameters (see below, footnote 4), while
the total transfer function $T(k)$ is taken from Ref.~\cite{Hu}.
The transfer function depends on $H_0$ and on baryon and cold dark matter density
parameters $\Omega_b$ and $\Omega_c$.
The total amount of matter is given by $\Omega_m = \Omega_b + \Omega_c$
and in this paper we take $\Omega_b h^2 = 0.02$, with $h$ defined by
\begin{equation}
\label{littleh}
H_0 = 100 h \; \km/\s/\Mpc \ .
\end{equation}
In the Press-Schechter parametrization, the form of the multiplicity function
is a result of the assumption that initial density fluctuations are Gaussian:
$\nu f(\nu) = e^{-\nu^2/2}/\sqrt{2\pi}$. In this paper, however, we use the form
\begin{equation}
\label{f}
\nu f(\nu) = K \! \left[ 1+(a\nu^2)^{-p} \right] \! e^{-a\nu^2/2} \ ,
\end{equation}
introduced by Sheth and Tormen, inspired by a model of elliptical collapse.
The constant
\begin{equation}
\label{K}
K = \frac{\sqrt{a}}{\sqrt{2\pi} + 2^{1/2 - p} \, \Gamma \! \left(1/2 - p \right)}
\end{equation}
results from the normalization condition~(\ref{intf}), $\Gamma(x)$ is the Gamma
function, while $a$ and $p$ are phenomenological constants to be determined by
fitting to $N$-body simulation results. We use the values found by Sheth and
Tormen, namely $a=0.707$ and $p=0.3$. (The Press-Schechter case is recovered for
$a=1$ and $p=0$.)

Finally, putting all this together, we can rewrite the mass function as
\begin{eqnarray}
\label{N2}
&& \!\!\!\!\!\!\! N(M>M_0,z) = \\
&& \!\!\!\!\!\!\! 2K \, \frac{\rho_m(z)}{M_0} \, \frac{\delta_c(z)}{\sigma_8 D(z)}
              \int_{1}^\infty \! \frac{dx}{x^3} \frac{1}{\Sigma^2(x)} \! \left| \frac{d\Sigma(x)}{dx} \right| \nonumber \\
&& \!\!\!\!\!\!\! \times \left\{ 1+\left[\frac{a\delta_c^2(z)}{\sigma_8^2 D^2(z)\Sigma^2(x)} \right]^{\!-p} \right\}
                  \exp \! \left[-\frac{a\delta_c^2(z)}{2\sigma_8^2 D^2(z)\Sigma^2(x)} \right] \ , \nonumber
\end{eqnarray}
where we have introduced the function
\begin{equation}
\label{Sigma}
\Sigma^2(x) = \frac{\int_0^\infty dy \, y^{n+2} \, T^2(y/R_8) \, W^2(xy R_0/R_8)}
{\int_0^\infty dy \, y^{n+2} \, T^2(y/R_8) \, W^2(y)} \ ,
\end{equation}
and the quantities\footnote{
As anticipated, the normalization constant $A$ can be related to
the other cosmological parameters, $\Omega_b$, $\Omega_c$, $h$, and $n$:
$A = 2 \pi^2 \sigma_8^2 R_8^{n+3} / \int_0^\infty dy \, y^{n+2} \, T^2(y/R_8) \, W^2(y)$.}
\begin{eqnarray}
\label{8}
&& \sigma_8 = \sigma(R_8,0) \ , \nonumber \\
&& R_8 = 8h^{-1} \Mpc \ , \\
&& R_0 = \left( \frac{3M_0}{4\pi\rho_m^{(0)}}\right)^{\!1/3} \ . \nonumber
\end{eqnarray}
We note that the function $\Sigma(x)$ evaluated at $x=R/R_0$ is the
present value of the rms amplitude $\sigma$ at the scale $R$
normalized to its present value at the scale $R_8$:
\begin{equation}
\label{sigmaSigma}
\Sigma(R/R_0) = \frac{\sigma(R,0)}{\sigma_8} \ .
\end{equation}
This result will be used in Appendix A.

\subsection{\normalsize{IIb. Data}}
\renewcommand{\thesection}{\arabic{section}}

Data on cluster abundance at different redshifts can be expressed as the comoving number
of clusters in the redshift interval $[z_1,z_2]$, with mass $M'$ within a reference
comoving radius $R_0'$ greater than a fiducial mass $M_0'$ within the same radius. Here
and in the following, a prime is used to distinguish quantities related to observed
masses and radii from theoretical ones, discussed in the previous subsection. We follow
Refs.\ \cite{Bahcall} and take
\begin{eqnarray}
\label{M0obs}
&&R_0' = 1.5 h^{-1}\Mpc \ ,
\\
&& M_0' = 8 \times 10^{14} h^{-1} M_\odot \ ,
\end{eqnarray}
where $M_\odot \simeq 1.989 \times 10^{33}$g is the solar mass.

Only an effective fraction $\alpha(z)$ of the total comoving volume at redshift $z$ is
observed, so the expected comoving number of clusters in the
redshift interval $[z_1,z_2]$, with mass $M'$ greater than $M_0'$, is
\begin{equation}
\label{NumberP}
\mathcal{N}' = \int_{z_1}^{z_2} \! dz \, \frac{d [\alpha(z) V(z)]}{dz} \, N'(M'>M_0',z) \ ,
\end{equation}
where $N'(M'>M_0',z)$ represents the comoving cluster number density
at redshift $z$ of clusters with masses $M'$ greater than $M_0'$.
The mass function $N'(M'>M_0',z)$ can be written as
\begin{equation}
\label{NP}
N'(M'>M_0',z) = \int_{M_0'}^\infty \! dM' n'(M',z) \ .
\end{equation}
Here $n'(M',z) \, dM'$ is the comoving number
density at redshift $z$ of clusters with masses $M'$ in the interval $[M',M'+dM']$, and
is defined by
\begin{equation}
\label{NNP}
n'(M',z) \, dM' = n(M,z) \, dM \ ,
\end{equation}
where $n(M,z)$ is given by Eq.~(\ref{n}).

Inserting Eq.~(\ref{NNP}) in Eq.~(\ref{NP}) we obtain
\begin{equation}
\label{N2P}
N'(M'>M_0',z) = \int_{g(M_0')}^\infty \! dM \, n(M,z) \ ,
\end{equation}
where the function $g$ relates the observed mass $M'$ to
the virial mass $M$ in the PS or ST parametrization. Consequently, $g(M_0')$ is the fiducial
virial mass
\begin{equation}
\label{vM}
M_0 = g(M_0')
\end{equation}
which corresponds to the fiducial mass $M_0'$ adopted in the observations.
In Appendix B we describe the procedure that gives the mass $M_0$ as a function of
$M_0'$. In general, the function $g$ depends on the redshift and cosmological parameters.

In Table~I, we show the four redshift bins $[z_1^{(i)},z_2^{(i)}]$ ($i=1,2,3,4$),
centered at $z_c^{(i)}$, of the massive clusters data. Also listed are the
values of the effective fraction of the observed comoving volume of each bin,
$\alpha_i$. The $\alpha_i$ values are for a cosmology with $(\Omega_m,w) = (0.3,-1)$,
and were computed using the results of Refs.\ \cite{Bahcall}.

The $\alpha_i$ parameters depend, in principle, on the cosmology and their values can be obtained
using the $\Sigma(1/V_{\rm max})$ method applied to observational data~\cite{Schmidt,Ikebe,Henry}.
However, the dependence of $\alpha_i$ on the cosmology is weak compared to that of the comoving
volume. Using the results of Refs.\ \cite{Bahcall} we get, for example, that passing from the
cosmology with $(\Omega_m,w) = (0.3,-1)$ to that with $\Omega_m = 1$, the percentage variation of
the comoving volume relative to the third bin, $V(z_2^{(3)}) - V(z_1^{(3)})$, is $117\%$,
while that of $\alpha_3$ is about $1\%$. Similar results hold for the other bins.


\begin{table}[t]
\vspace*{-0.3cm}
\caption{The four redshift intervals $[z_1^{(i)},z_2^{(i)}]$ ($i=1,2,3,4$),
centered at $z_c^{(i)}$, of the massive clusters data
and the references from which data have been taken. $\alpha_i$ is the
effective fraction of the observed comoving volume of the
$i^{\rm th}$ bin and the values listed here are for the case
with $\Omega_m = 0.3$ and $w=-1$.}

\vspace{0.5cm}

\begin{tabular}{ccccccc}

\hline \hline

&bin $i$  &$z_1^{(i)}$  &$z_2^{(i)}$  &$z_c^{(i)}$  &Ref.  &$\alpha_i$ \\

\hline

&1  &0.00  &0.10 &0.050 &\cite{Ikebe}    &0.309 \\
&2  &0.30  &0.50 &0.375 &\cite{Henry}    &0.012 \\
&3  &0.50  &0.65 &0.550 &\cite{Bahcall}  &0.006 \\
&4  &0.65  &0.90 &0.825 &\cite{Donahue2} &0.001 \\

\hline \hline

\end{tabular}
\end{table}


References \cite{Bahcall,Ikebe,Henry,Donahue2} give X-ray temperature measurements for massive
clusters. For completeness we show these data in Table~II.


\begin{table}[t]
\vspace*{-0.2cm}
\caption{Name and X-ray temperature $T_X$
of clusters in the four bins used in our analysis.
Data in the first and second bins are from Ref.~\cite{Ikebe} and Ref.~\cite{Henry},
respectively, while data for the third and fourth bins are
from Ref.~\cite{Bahcall} and Ref.~\cite{Donahue2}, respectively.
All errors are at the $68 \%$ confidence level.}

\vspace{0.5cm}

\begin{tabular}{lllc}

\hline \hline

&bin $i$ \ \  &name    &$T_X(\keV)$ \\

\hline

&   &A0754        &$9.00^{+0.21}_{-0.21}$ \\
&   &A2142        &$8.46^{+0.32}_{-0.30}$ \\
&   &COMA         &$8.07^{+0.18}_{-0.16}$ \\
&   &A2029        &$7.93^{+0.24}_{-0.22}$ \\
&   &A3266        &$7.72^{+0.21}_{-0.17}$ \\
&   &A0401        &$7.19^{+0.17}_{-0.15}$ \\
&   &A0478        &$6.91^{+0.24}_{-0.22}$ \\
&   &A2256        &$6.83^{+0.14}_{-0.13}$ \\
&   &A3571        &$6.80^{+0.13}_{-0.11}$ \\
&   &A0085        &$6.51^{+0.10}_{-0.14}$ \\
&   &A0399        &$6.46^{+0.23}_{-0.22}$ \\
&   &ZwCl1215     &$6.36^{+1.79}_{-1.23}$ \\
&1  &A3667        &$6.28^{+0.16}_{-0.16}$ \\
&   &A1651        &$6.22^{+0.27}_{-0.25}$ \\
&   &A1795        &$6.17^{+0.16}_{-0.15}$ \\
&   &A2255        &$5.92^{+0.24}_{-0.16}$ \\
&   &A3391        &$5.89^{+0.27}_{-0.20}$ \\
&   &A2244        &$5.77^{+0.37}_{-0.27}$ \\
&   &A0119        &$5.69^{+0.15}_{-0.17}$ \\
&   &A1650        &$5.68^{+0.18}_{-0.16}$ \\
&   &A3395s       &$5.55^{+0.54}_{-0.40}$ \\
&   &A3158        &$5.41^{+0.16}_{-0.15}$ \\
&   &A2065        &$5.37^{+0.21}_{-0.18}$ \\
&   &A3558        &$5.37^{+0.10}_{-0.09}$ \\
&   &A3112        &$4.72^{+0.23}_{-0.15}$ \\
&   &A1644        &$4.70^{+0.55}_{-0.43}$ \\

\hline

&   &MS 1008.1    &$8.2^{+1.2}_{-1.1}$   \\
&   &MS 1358.4    &$6.9^{+0.5}_{-0.5}$   \\
&2  &MS 1621.5    &$6.6^{+0.9}_{-0.8}$   \\
&   &MS 0353.6    &$6.5^{+1.0}_{-0.8}$   \\
&   &MS 1426.4    &$6.4^{+1.0}_{-1.2}$   \\
&   &MS 1147.3    &$6.0^{+1.0}_{-0.7}$   \\

\hline

&3  &MS 0451--03  &$10.4^{+0.7}_{-0.7}$ \\
&   &MS 0016+16   &$8^{+0.6}_{-0.6}$        \\

\hline

&4  &MS 1054--03  &$12.3^{+1.9}_{-1.3}$ \\

\hline \hline

\end{tabular}
\end{table}



\begin{table*}[t]
\caption{The threshold X-ray temperature, $T_{X,0}$, and the observed number
of clusters with masses $M' > M'_0$, $\mathcal{N}'_{{\rm obs},i}$,
in each bin and for each $\Delta_v'$ interval.
The uncertainty on the comoving numbers of clusters,
$\Delta {\mathcal{N}'_{{\rm obs},i}}$, is also indicated.}

\vspace{0.5cm}

\begin{tabular}{llcccc}

\hline \hline

&   &bin 1  &bin 2  &bin 3  &bin 4 \\

&   &$[T_{X,0}(\keV) \, , \mathcal{N}'_{{\rm obs},1}]$  &$[T_{X,0}(\keV) \, , \mathcal{N}'_{{\rm obs},2}]$
    &$[T_{X,0}(\keV) \, , \mathcal{N}'_{{\rm obs},3}]$  &$[T_{X,0}(\keV) \, , \mathcal{N}'_{{\rm obs},4}]$ \\

\hline

&$\Delta_v' \in    [25,175]$     &$[7.37\, , 5^{+1}_{-0}]$  &$[9.6\, ,0^{+0}_{-0}]$  &$[10.9\, ,0^{+1}_{-0}]$  &[$12.8\, ,0^{+1}_{-0}]$  \\
&$\Delta_v' \in \; ]175,375]$    &$[6.15\, ,15^{+2}_{-4}]$  &$[8.1\, ,1^{+0}_{-1}]$  &$[9.1 \, ,1^{+1}_{-0}]$  &[$10.7\, ,1^{+0}_{-0}]$  \\
&$\Delta_v' \in \; ]375,750]$    &$[5.54\, ,21^{+2}_{-5}]$  &$[7.3\, ,1^{+4}_{-1}]$  &$[8.2 \, ,2^{+0}_{-1}]$  &[$9.6 \, ,1^{+0}_{-0}]$  \\
&$\Delta_v' \in \; ]750,1750]$   &$[5.14\, ,24^{+1}_{-1}]$  &$[6.7\, ,2^{+4}_{-1}]$  &$[7.6 \, ,2^{+0}_{-1}]$  &[$8.9 \, ,1^{+0}_{-0}]$  \\
&$\Delta_v' \in \; ]1750,3250]$  &$[4.91\, ,24^{+2}_{-0}]$  &$[6.4\, ,5^{+1}_{-3}]$  &$[7.3 \, ,2^{+0}_{-0}]$  &[$8.5 \, ,1^{+0}_{-0}]$  \\

\hline \hline

\end{tabular}
\end{table*}


In order to convert temperature to mass, we use the mass-temperature conversion formula of
Ref.\ \cite{Hjorth} (see Ref.\ \cite{T1} for a different approach to the problem of
cluster mass-temperature conversion):
\begin{equation}
\label{MT1}
M(< r) = 10^{14} M_\odot \: \kappa_\Delta \, \frac{T_X}{\keV} \, \frac{r}{\Mpc} \, \ ,
\end{equation}
where $M(< r)$ is the mass within a physical radius $r$,
$T_X$ is the cluster X-ray temperature,
and $\kappa_\Delta$ is a parameter which depends only on $\Delta'_v$.
Here $\Delta'_v$ is the virial overdensity relative to the critical density.
It is related to the virial overdensity relative to the background matter density,
$\Delta_v$, through
\begin{equation}
\label{DeltaDelta}
\Delta'_v = \frac{\Omega_m (1+z)^3}{E^2(z)} \, \Delta_v \ .
\end{equation}
The quantity $\Delta_v$ depends on redshift and cosmology and is thoroughly
discussed in Appendix C.
\footnote{
Defining $r_\Delta$ as the physical radius containing an overdensity of
$\Delta_v'$ relative to the critical density, and $M(r < r_\Delta)$ as the
mass contained in $r_\Delta$, the mass-temperature relation~(\ref{MT1})
assumes the standard $T_X^{3/2}$ power-law form~\cite{Hjorth}, namely
$M(r < r_\Delta) = \kappa_\Delta^{3/2} \, \left[3/4\pi \Delta'_v
\rho_{\rm cr}^{(0)}\right]^{1/2} \, (1+z)^{-3/2} \, T_X^{3/2}$.}
As found in Ref.~\cite{Hjorth}, the parameter $\kappa_\Delta$ depends on $\Delta_v'$
and, in particular, when
\begin{equation}
\label{Deltanumbers}
\Delta'_v = 100, 250, 500, 1000, 2500 \ ,
\end{equation}
$\kappa_\Delta$ assumes, respectively, the values
\begin{equation}
\label{kappanumbers}
\kappa_\Delta = 0.76, 0.91, 1.01, 1.09, 1.14 \ .
\end{equation}
From Eq.~(\ref{MT1}), it follows that a mass $M'$ within a comoving radius $R_0' = 1.5 h^{-1}\Mpc$
is related to the X-ray temperature by
\begin{equation}
\label{MT2}
M' = 1.5 \times 10^{14} h^{-1} M_\odot \: \kappa_\Delta \, \frac{T_X}{\keV} \, \frac{1}{1+z} \ .
\end{equation}
This means that clusters in the $i^{\rm th}$ bin, with masses $M' > M_0$,
will have a temperature exceeding the threshold value
\begin{equation}
\label{threshold}
\frac{T_{X,0}}{\keV} = \frac{16 (1+z_c^{(i)})}{3\kappa_\Delta} \ .
\end{equation}
In order to apply the above equation to data, we extrapolate the parameter
$\kappa_\Delta$ according to the following prescription:
\begin{equation}
\label{prescription}
\kappa_\Delta =
\left\{ \begin{array}{ll}
   0.76 & \; \mbox{if} \;\; \Delta_v' \in    [25,175] \ ,
\\
   0.91 & \; \mbox{if} \;\; \Delta_v' \in \; ]175,375] \ ,
\\
   1.01 & \; \mbox{if} \;\; \Delta_v' \in \; ]375,750] \ ,
\\
   1.09 & \; \mbox{if} \;\; \Delta_v' \in \; ]750,1750] \ ,
\\
   1.14 & \; \mbox{if} \;\; \Delta_v' \in \; ]1750,3250] \ .
\end{array}
\right.
\end{equation}
By using Eqs.\ (\ref{threshold}) and (\ref{prescription}) and data from Table~II we find
the values listed in Table~III for the observed number of clusters with masses $M' > M'_0$ in the
$i^{\rm th}$ bin, $\mathcal{N}'_{{\rm obs},i}$.

The uncertainty in the comoving numbers of clusters,
$\Delta {\mathcal{N}'_{{\rm obs},i}}$, derive from the uncertainty
in the X-ray temperature of clusters. The threshold X-ray temperature
in each bin and for each $\Delta_v'$ interval is also indicated in
Table~III.

Finally, we calculate the observed number of clusters in the four redshift bins as
\begin{equation}
\label{NExpected}
\mathcal{N}'_i = \alpha_i \int_{z_1^{(i)}}^{z_2^{(i)}} \! dz \, \frac{dV(z)}{dz} \, N'(M'>M_0',z) \ ,
\end{equation}
where we used Eq.~(\ref{NumberP}) and replaced $\alpha(z)$ with the
average effective fraction $\alpha_i$ in the $i^{\rm th}$ bin. Since, as argued above,
the $\alpha_i$ only depend weakly on the cosmology, we use in Eq.~(\ref{NExpected}), for definiteness,
the values for the case $\Omega_m = 0.3$ and $w=-1$ listed in Table~I.

In Fig.~1 we plot the various quantities needed in the computation of the
mass function as a function of redshift and for different choices of $(w_0,w_a)$.
We fixed the values of the other cosmological parameters to the best-fit values
obtained by using the 7-year WMAP observations~\cite{WMAP7}, namely
$(h,n,\Omega_m,\sigma_8) = (0.71,0.96,0.3,0.80)$. Although the comoving volume
is very sensitive to the choice of the cosmological model, the variations of the
functions $\delta_c$, $D(z)$, $M_0/M_0'$, and $\Delta_v$ (discussed in the Appendices)
with redshift and cosmology concur to give rise to larger changes in
$N'(M'>M_0',z)$, especially at large redshifts.
This, in principle, can be used to put constraints on various models of dark energy.


\begin{figure*}[t]
\vspace*{-1.7cm}
\begin{center}
\hspace*{-2.3cm}
\includegraphics[clip,width=1.25\textwidth]{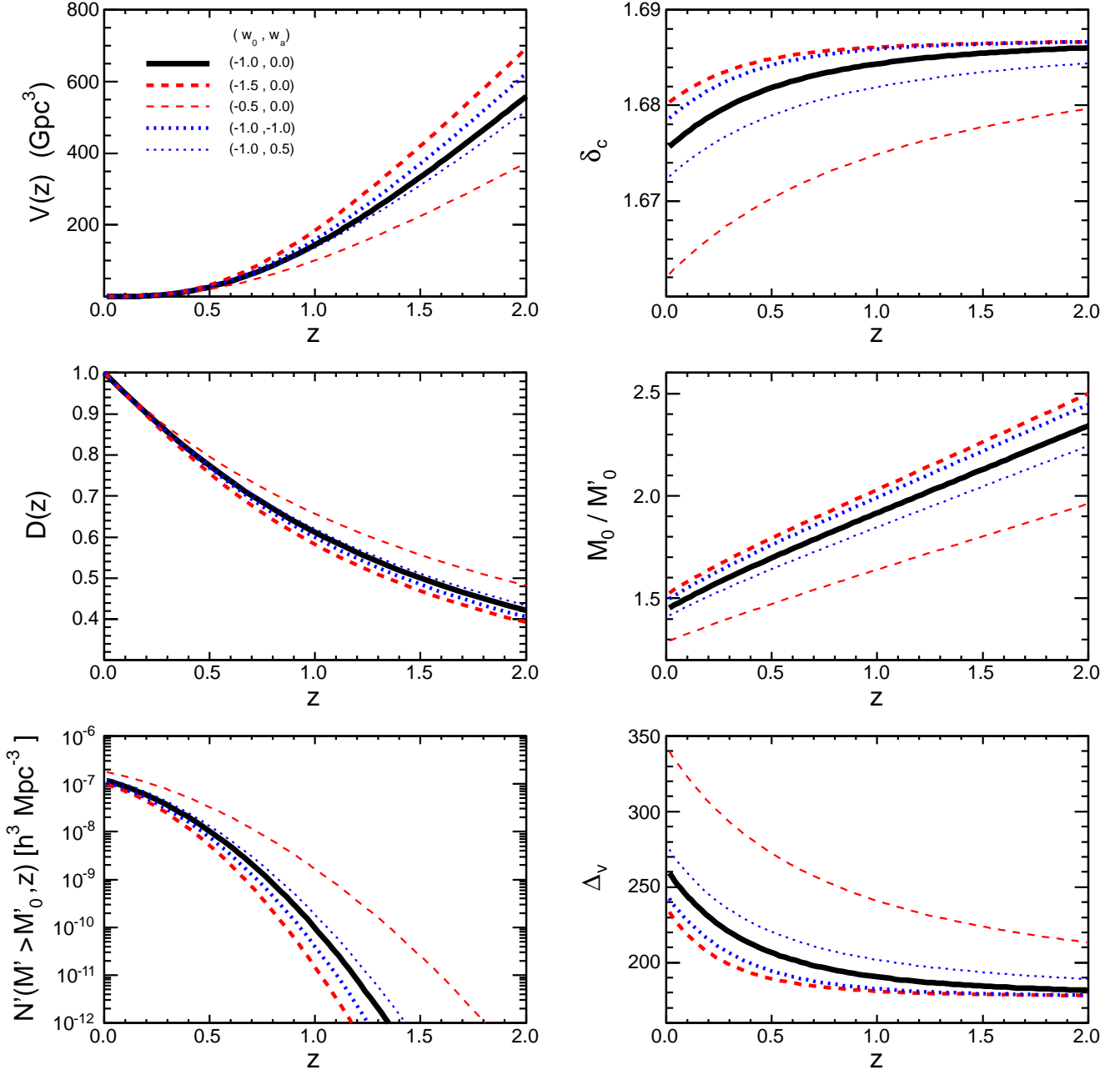}
\vspace*{-2cm}
\caption{The comoving volume $V(z)$, the critical density contrast $\delta_c(z)$,
the growth factor $D(z)$, the mass ratio $M_0/M_0'$, the mass function $N'(M'>M_0',z)$
[see Eq.~(\ref{N2P})], and the virial overdensity relative to the background matter density,
$\Delta_v$, as a function of the redshift $z$, for different dark energy
equation of state parameters $(w_0,w_a)$. We fixed the values of the other cosmological parameters to
$(h,n,\Omega_m,\sigma_8) = (0.71,0.96,0.3,0.80)$.} 
\end{center}
\end{figure*}


Due to the small number of clusters in each bin, the comparison between observed
and predicted number of clusters is made using Poisson error statistics. Accordingly,
we define a likelihood function by
\begin{equation}
\label{likelihood}
\mathcal{L} = \prod_{i=1}^4 \frac{\lambda_i^{\kappa_i} e^{-\lambda_i}}{\kappa_i!} \, ,
\end{equation}
where we have introduced $\lambda_i \equiv \mathcal{N}'_i$ and $\kappa_i \equiv {\mathcal{N}'_{{\rm obs},i}}$
for notational clarity.  The $\chi^2$ statistics is then introduced as
\begin{eqnarray}
\label{chi2N-nosys}
&& \!\!\!\!\!
\chi^2_{\rm CL, no \, sys}(h,n,\Omega_m,\sigma_8,w_0,w_a) =  -2\ln \mathcal{L} \\
&& \simeq 2\sum_{i=1}^4 \left[ \mathcal{N}'_i - {\mathcal{N}'_{{\rm obs},i}}
\left(1 + \ln \mathcal{N}'_i - \ln {\mathcal{N}'_{{\rm obs},i}} \right) \right] \ . \nonumber
\end{eqnarray}
We also take into account the uncertainty in the comoving numbers of clusters,
$\Delta {\mathcal{N}'_{{\rm obs},i}}$, by shifting the observed number of clusters in each bin as
\begin{equation}
\label{antony}
{\mathcal{N}'_{{\rm obs},i}}  \rightarrow  {\mathcal{N}'_{{\rm obs},i}} + \xi \Delta {\mathcal{N}'_{{\rm obs},i}}
\equiv {\mathcal{N}''_{{\rm obs},i}}
\end{equation}
where the ``pull'' $\xi$ is a univariate gaussian random variable~\cite{Getting}. Correspondingly, we modify the
$\chi^2$ as
\begin{eqnarray}
\label{chi2N}
&& \!\!\!\!\!
\chi^2_{\rm CL}(h,n,\Omega_m,\sigma_8,w_0,w_a,\xi) = \\
&& = 2\sum_{i=1}^4 \left[ \mathcal{N}'_i - {\mathcal{N}''_{{\rm obs},i}}
\left(1 + \ln \mathcal{N}'_i - \ln {\mathcal{N}''_{{\rm obs},i}} \right) \right] + \xi^2 \ . \nonumber
\end{eqnarray}


\section{\normalsize{III. Other Cosmological Data}}
\renewcommand{\thesection}{\arabic{section}}

In this section, we present data from other type of cosmological observations. In the
next section, we derive joint constraints using these data along with those of massive
clusters.

{\it Baryon Acoustic Oscillations.}-- The measurement of baryon acoustic oscillations
(BAOs) in the large-scale matter correlation function, fixes the values of a
characteristic ``BAO distance parameter'', which we denote as $\mathcal{C}$.\footnote{
See Refs.\ \cite{Gaztanagaetal} for recent discussions of BAO data constraints on cosmological
parameters.}

With $D_V$ an effective distance defined by
\begin{equation}
\label{DV}
D_V(z) =  \frac{1}{H_0} \left( \frac{z}{E(z)} \right)^{\!\!1/3} \,
          \left[\int_0^{z} \! \frac{dz'}{E(z')} \right]^{2/3} \ ,
\end{equation}
the $\mathcal{C}$ parameter is the ratio
\begin{equation}
\label{C}
\mathcal{C} =  \frac{r_s(z_d)}{D_V(0.275)}
\end{equation}
between the comoving sound horizon at the baryon drag epoch $z_d$,
\begin{equation}
\label{rs}
r_s(z) = \frac{1}{\sqrt{3} \, H_0} \int_z^{\infty} \! \frac{dz'}{E(z')} \,
         \sqrt{\frac{4\Omega_\gamma (1+z)}{4\Omega_\gamma (1+z) + 3\Omega_b}} \ ,
\end{equation}
and the effective distance $D_V$ at $z=0.275$~\cite{Percival}. Here,
$\Omega_\gamma$ is the photon density parameter that we take equal to
$\Omega_\gamma h^2 = 2.56 \times 10^{-5}$~\cite{Kolb}.
\footnote{Since the upper limit of integration of the integral
in Eq.~(\ref{rs}) is infinity we must include, in the expression of the
normalized Hubble parameter $E(z)$, the contribution due to
radiation. Accordingly, we add the quantity
$\rho_r/\rho_{\rm cr}^{(0)} = \Omega_r (1+z)^4$
in the argument of the square root appearing in Eq.~(\ref{H}),
where $\rho_r$ and $\Omega_r$ are the radiation energy density and
radiation density parameter, respectively. We take
$\Omega_r h^2 = 4.31 \times 10^{-5}$~\cite{Kolb}.}
For the redshift at the baryon drag epoch, $z_d$, we use the fitting
formula of Ref.\ \cite{Hu}:
\begin{equation}
\label{zd}
z_d = \frac{1291 (\Omega_m h^2)^{0.251}}{1+ 0.659 (\Omega_m h^2)^{0.828}} \left[ 1+ b_1 (\Omega_b h^2)^{b_2} \right] \ ,
\end{equation}
where
\begin{eqnarray}
\label{b1b2}
&& \!\!\!\!\!\!\!\!\!\!\!\!\!\! b_1 = 0.313 (\Omega_m h^2)^{-0.419} \left[ 1 + 0.607 (\Omega_m h^2)^{0.674} \right] \ , \\
&& \!\!\!\!\!\!\!\!\!\!\!\!\!\! b_2 = 0.238 (\Omega_m h^2)^{0.223} \ .
\end{eqnarray}
BAO data give the value $\mathcal{C}_{\rm obs} = 0.1390 \pm 0.0037$~\cite{Percival}.
Accordingly, we define the $\chi^2$ statistic
\begin{equation}
\label{chi2C}
\chi^2_{\rm BAO}(h,\Omega_m,w_0,w_a) =
\frac{(\mathcal{C} - 0.1390)^2}{0.0037^2} \ .
\end{equation}

{\it Cosmic Microwave Background.}-- The analysis of the CMB radiation puts a constraint
on the reduced distance to the surface of last scattering, the so-called ``CMB shift parameter'',
\begin{equation}
\label{R}
\mathcal{R} = \Omega_m^{1/2} \int_0^{z_{\rm ls}} \! \frac{dz}{E(z)} \ ,
\end{equation}
where $z_{\rm ls} \simeq 1090$ is the redshift at the time of last scattering.
The shift parameter weakly depends on the adopted cosmology and here we
use the constraint found by Corasaniti and Melchiorri,
$\mathcal{R}_{\rm obs} = 1.710 \pm 0.026$~\cite{Melchiorri},
which refers to a cosmological model with evolving dark energy with equation-of-state
parameter of the form given in Eq.~(\ref{wz}).
We then consider the following $\chi^2$ statistic
\begin{equation}
\label{chi2R}
\chi^2_{\rm CMB}(\Omega_m,w_0,w_a) =
\frac{\left( \mathcal{R} - 1.710 \right)^2}{0.026^2} \ .
\end{equation}

{\it Hubble Constant.}-- A meta-analysis of many measurements yields
$H_0 = (68 \pm 2.8) \, \km/\s/\Mpc$ at $1\sigma \; \mbox{C.L.}$
\cite{Ratra}.\footnote{
This is reasonably consistent with both `low' \cite{Tammann} and `high'
\cite{Freedman} recent estimates of the Hubble constant.}
Accordingly, we introduce the penalty
\begin{equation}
\label{chi2H0}
\chi^2_h(h) =
\frac{(h - 0.68)^2}{0.028^2} \ .
\end{equation}

{\it Hubble Parameter.}-- The analysis of spectra of passive\-ly-evolving red galaxies
enables the determination of the Hubble parameter at different redshifts, \cite{Jimenez}.
We use data quoted in Ref.~\cite{Stern} and reported in Table~IV for the sake of completeness.
To these data we also add the estimate of the Hubble parameter at redshifts $z=0.24$ and
$z=0.43$, obtained in Ref.~\cite{Gaztanaga} by using the BAO peak position as a standard ruler
in the radial direction.\footnote{
See Refs.\ \cite{hubbleparameter} for Hubble parameter measurement constraints on cosmological
parameters.}

We then introduce a $\chi^2$ statistic as
\begin{equation}
\label{chi2H}
\chi^2_{\rm Hubble}(h,\Omega_m,w_0,w_a) =
\sum_{i=1}^{13} \frac{\left[ H(z_i) - H_{\rm obs}(z_i) \right]^2}{\sigma_H^2} \ .
\end{equation}


\begin{table}[t]
\vspace*{-0.2cm}
\caption{The observed Hubble parameter $H_{\rm obs}(z_i)$ with error $\sigma_H$
(in brackets) from passively evolving galaxies (data from Ref.~\cite{Stern}) and line-of-sight BAO
peak position (data are from Ref.~\cite{Gaztanaga} and marked with an asterisk).}

\vspace{0.5cm}

\begin{tabular}{ccc}

\hline \hline

&$z_i$        &$~~~H_{\rm obs}(z_i) \;\, [\,\km/\s/\Mpc\,]$ \\

\hline

&$0.1$  &$69(12)$  \\
&$0.17$ &$83(8)$   \\
&$0.24$ &$79.69(2.65)^*$  \\
&$0.27$ &$77(14)$  \\
&$0.4$  &$95(17)$  \\
&$0.43$ &$86.45(3.68)^*$  \\
&$0.48$ &$97(60)$  \\
&$0.88$ &$90(40)$  \\
&$0.9$  &$117(23)$ \\
&$1.3$  &$168(17)$ \\
&$1.43$ &$177(18)$ \\
&$1.53$ &$140(14)$ \\
&$1.75$ &$202(40)$ \\

\hline \hline

\end{tabular}
\end{table}



\begin{figure*}[t]
\vspace*{-0.5cm}
\begin{center}
\hspace*{-0.9cm}
\includegraphics[clip,width=1.04\textwidth]{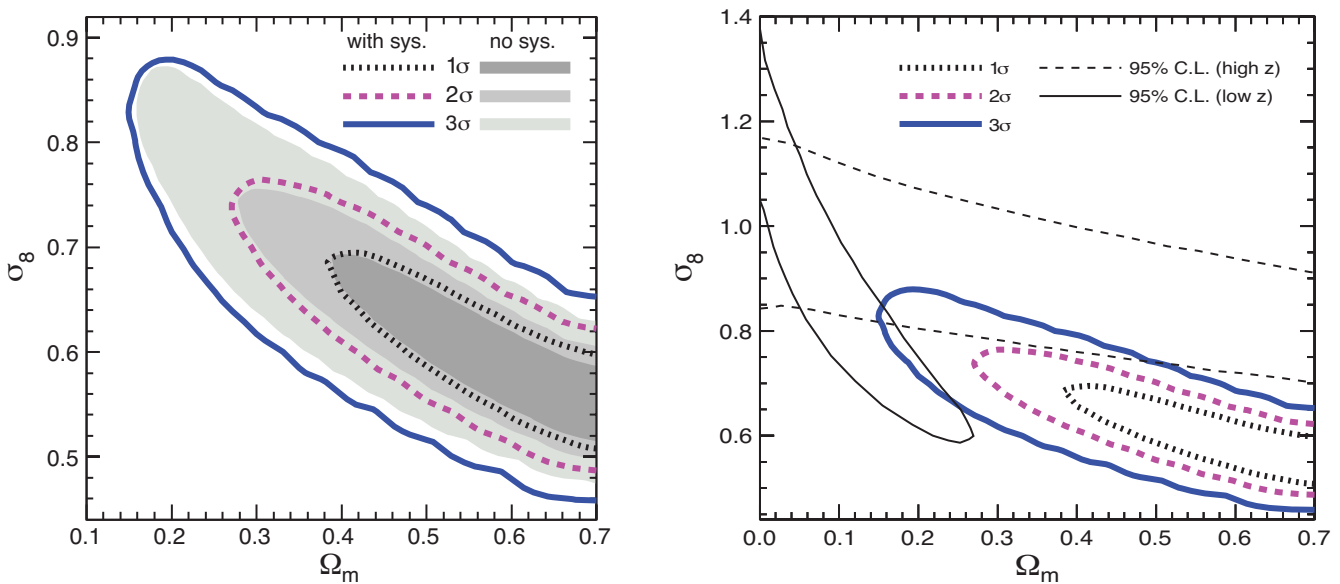}
\vspace*{-1.3cm}
\caption{{\it Left panel.} $1\sigma$, $2\sigma$, and $3\sigma$ confidence level contours in the
$(\Omega_m,\sigma_8)$ plane from galaxy cluster number count data. Results obtained by including
systematic uncertainties are shown as empty contours, while those ignoring systematics
(i.e., keeping just best fit values for data) are represented as filled contours.
{\it Right panel.} Thick contours are the $1\sigma$, $2\sigma$, and $3\sigma$
($\Delta \chi^2 = 1,4,9$) confidence level contours in the $(\Omega_m,\sigma_8)$ plane from
galaxy cluster number count data (including systematic errors; the same as in the left panel).
Thin continuous and thin dashed contours are the $95\%$ contour levels
(2 d.o.f., $\Delta \chi^2 = 5.99$) for the first two redshift bins (low $z$) and last
two redshift bins (high $z$), respectively, graphically reprodued from Refs.\ \cite{Bahcall}.}
\end{center}
\end{figure*}



\begin{figure}[t]
\vspace*{-0.3cm}
\begin{center}
\hspace*{-0.8cm}
\includegraphics[clip,width=0.52\textwidth]{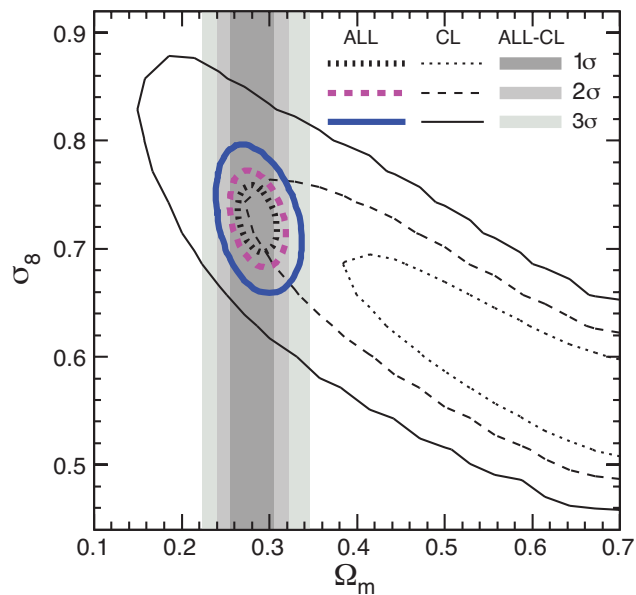}
\vspace*{-1.2cm}
\caption{$1\sigma$, $2\sigma$, and $3\sigma$ confidence level contours in the
$(\Omega_m,\sigma_8)$ plane from galaxy cluster number count (including  systematic errors),
BAO, CMB, Hubble parameter, and SNe observations. Filled vertical bands are the result of
combining BAO, CMB, Hubble parameter, and SN data (ALL-CL), empty open thin contours give the
confidence level contours for cluster data only (CL), while empty closed thick contours are
from the combination of all data (ALL).}
\end{center}
\end{figure}



\begin{figure*}[t]
\vspace*{-0.5cm}
\begin{center}
\hspace*{-0.9cm}
\includegraphics[clip,width=1.04\textwidth]{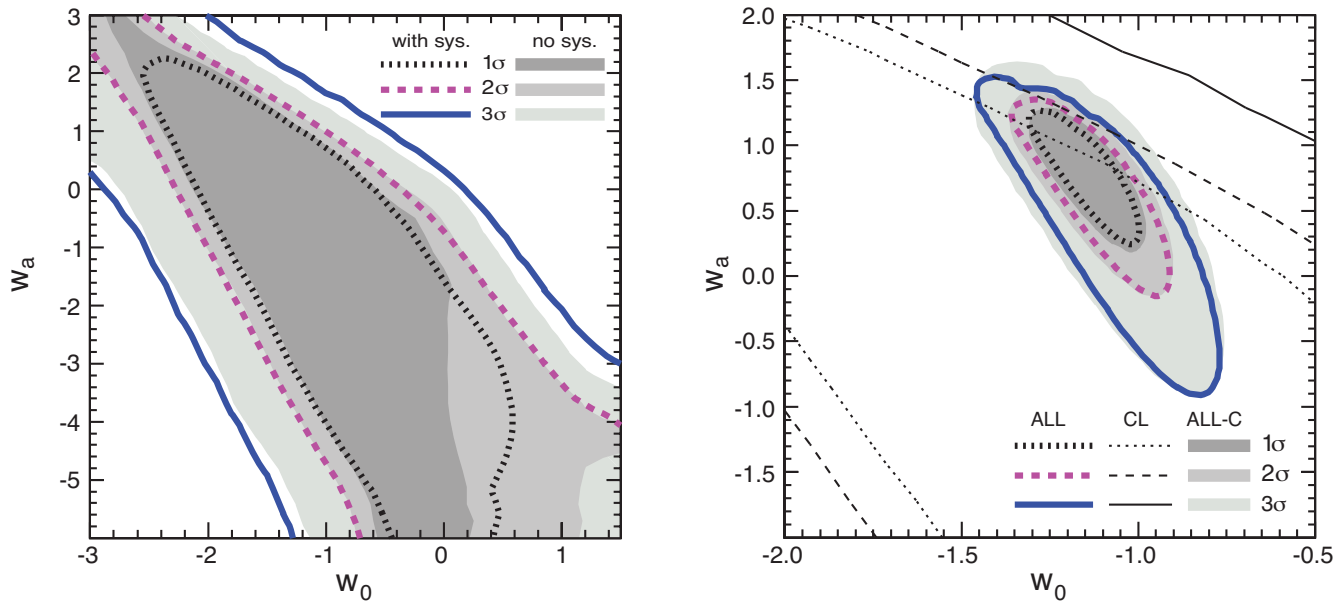}
\vspace*{-1.3cm}
\caption{$1\sigma$, $2\sigma$, and $3\sigma$ confidence level contours in the
$(w_0,w_a)$ plane. {\it Left panel.} Results from galaxy cluster number count data
obtained by including systematic uncertainties are shown as empty contours, while those
ignoring systematics (i.e., keeping just best fit values for data) are represented as filled
contours. {\it Right panel.} Results determined from galaxy cluster number count
(including  systematic errors), BAO, CMB, Hubble parameter, and SNe observations.
Filled contours are the result of combining BAO, CMB, Hubble parameter, and SN data (ALL-CL),
empty open thin contours give the confidence level contours for cluster data only (CL),
while empty closed thick contours are from the combination of all data (ALL).}
\end{center}
\end{figure*}



\begin{figure*}[t]
\vspace*{-0.5cm}
\begin{center}
\hspace*{-0.9cm}
\includegraphics[clip,width=1.04\textwidth]{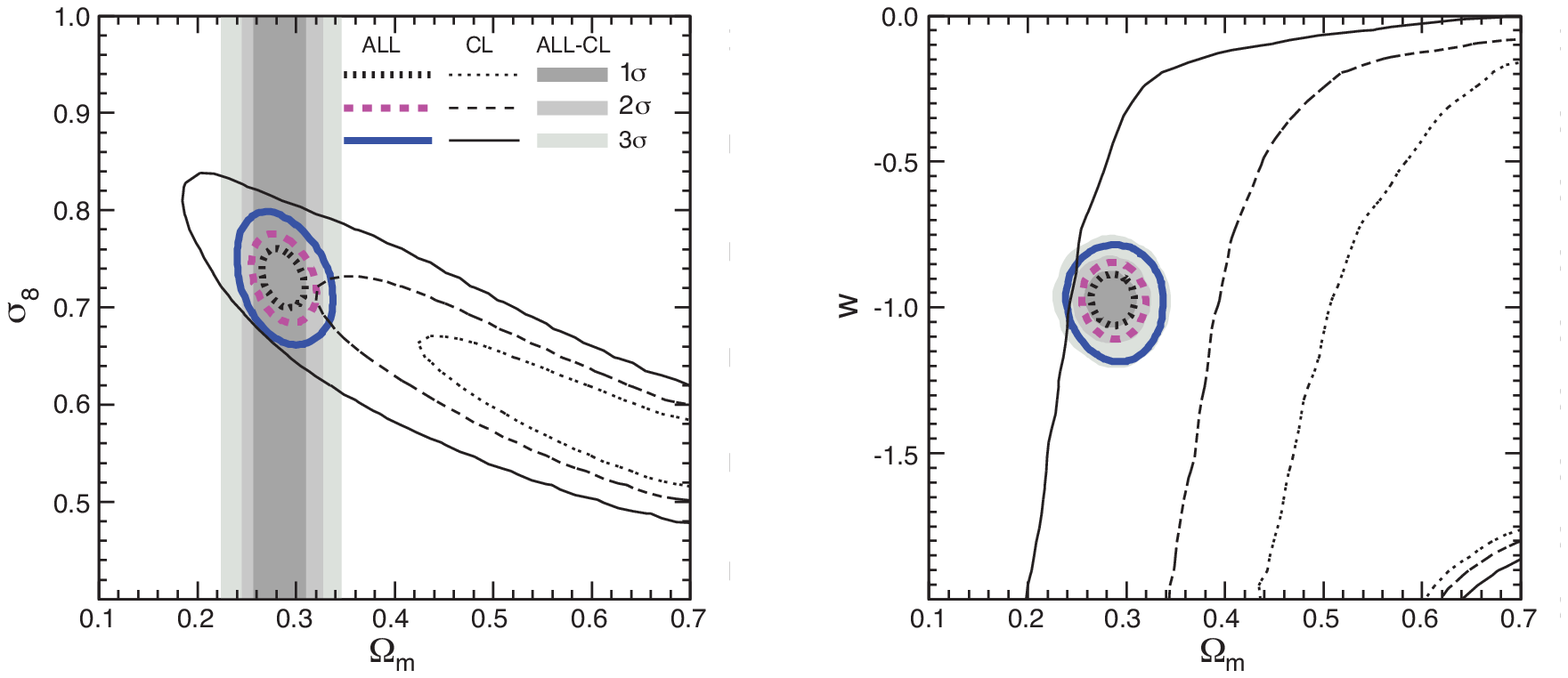}
\vspace*{-1.3cm}
\caption{$1\sigma$, $2\sigma$, and $3\sigma$ confidence level contours in the
$(\Omega_m,\sigma_8)$ plane (left panel) and in the $(\Omega_m,w)$ plane (right panel), for
the XCDM parametrization, determined from galaxy cluster number count
(including systematic errors), BAO, CMB, Hubble parameter, and SNe observations.
Filled contours are the result of combining BAO, CMB, Hubble parameter, and SN data (ALL-CL),
empty open thin contours give the confidence level contours for cluster data only (CL),
while empty closed thick contours come from the combination of all data (ALL).}
\end{center}
\end{figure*}


{\it Type Ia supernovae.}-- Type Ia supernovae are standardizable candles and
so can be used to discriminate between different cosmological models. Indeed,
the theoretically-predicted distance modulus $\mu$, defined by
\begin{equation}
\label{mu} \mu(z) = 5 \, \log_{10} \! \left( \frac{d_L}{1\Mpc} \right) + 25  \ ,
\end{equation}
depends on the redshift and on the set of cosmological parameters $(h,\Omega_m,w_0,w_a)$
and can be compared to the one ``derived'' from the observation
of SN lightcurves~\cite{Union2}, namely $\mu_B$. This, in turn,
is deduced from the analysis of SN lightcurves which, if performed
using the ``SALT2'' fitter~\cite{SALT2},
gives~\cite{Union2}
\begin{equation}
\label{new8} \mu_B = m_B^{\rm max} - M + \alpha x_1 - \beta c\ .
\end{equation}
Here, $m_B^{\rm max}$ and $c$ are the peak bolometric apparent magnitude and the color correction,
respectively, while $x_1$ is a SALT2 fitter parameter~\cite{SALT2}.
The absolute magnitude of SNe, $M$, and $\alpha$ and $\beta$
are, instead, nuisance parameters to be determined, simultaneously with
the cosmological parameters from fits to data. In this paper, we use
data from the Union2 SN compliation \cite{Union2} which
consists of 557 SNe.

However, since the covariance matrices resulting from the lightcurve fit
are not publicly available, we do not have any information on the correlation between the errors
on $m_B^{\rm max}$, $x_1$, and $c$. Consequently, we follow the analysis
of Ref.~\cite{Lewis,Melchiorri} as explained in Ref.~\cite{Marrone}
and introduce the $\chi^2$ statistic
\begin{eqnarray}
\label{new10}
&& \!\!\!\!\!\!\!\!\!\!\!\!\!\!\!\!
\chi^2_{\rm SN}(\Omega_m,w_0,w_a) \nonumber \\
&& = \sum_{ij} \left( \mu_i^{\rm exp} - \tilde{\mu}_i \right) \left( \sigma^{-2}_{ij} - M_{ij} \right)
\left( \mu_j^{\rm exp} - \tilde{\mu}_j \right)\ .
\end{eqnarray}
The double sum runs over the 557 SNe, $\mu_i^{\rm exp}$ is the experimental value of
the distance modulus of the $i^{\rm th}$ supernova, and
\begin{equation}
\label{new12} \tilde{\mu}_i = 5 \, \log_{10} \tilde{d}_L  + 25 \ ,
\end{equation}
is the ``reduced'' theoretical distance modulus. The ``reduced'' luminosity distance $\tilde{d}_L$ is
\begin{equation}
\label{new13} \tilde{d}_L = H_0 d_L \ ,
\end{equation}
and $\sigma^{2}_{ij}$ is the covariance matrix (containing both statistical and systematic errors), while
the matrix $M_{ij}$ is given by
\begin{equation}
\label{new11} M_{ij} = \frac{\sum_{kl} \sigma^{-2}_{ik} \sigma^{-2}_{lj}}{\sum_{kl} \sigma^{-2}_{kl}} \ .
\end{equation}
It is worth noting that $\tilde{d}_L$ is independent of
the Hubble parameter $H_0$, so the $\chi^2$ in Eq.~(\ref{new10}) depends only
on the cosmological parameters $(\Omega_m,w_0,w_a)$.


There are many other data sets that can be used to constrain cosmological parameters,
for example, strong gravitational lensing observations \cite{GL}; however, for our
illustrative purposes here the data described above suffice.


\section{\normalsize{IV. Combined Data Analysis}}
\renewcommand{\thesection}{\arabic{section}}

In this section, we present the results of a joint analysis of massive cluster evolution,
BAO peak length, CMB anisotropy, Hubble parameter, and SNe apparent magnitude data.
The $\chi^2$ statistic is
\begin{eqnarray}
\label{chi2}
&& \!\!\!\!\!\!\!\!\!
\chi^2(h,n,\Omega_m,\sigma_8,w_0,w_a,\xi) \\
&& = \chi^2_{\rm CL} + \chi^2_{\rm BAO} + \chi^2_{\rm CMB} + \chi^2_h + \chi^2_{\rm Hubble} + \chi^2_{\rm SN} \ , \nonumber
\end{eqnarray}
and depends on the six cosmological parameters
$(h,n,\Omega_m,\sigma_8,w_0,w_a)$, and on the pull $\xi$.

Since the $\chi^2$ depends on seven parameters, a grid-based analysis is not feasible and
we therefore employ a Markov Chain Monte Carlo approach. We use a modified version
of CosmoMC~\cite{Lewis} to produce and analyze the likelihood chains.

\subsection{\normalsize{IVa. ($\Omega_m$,$\sigma_8$) results}}
\renewcommand{\thesection}{\arabic{section}}

In the left panel of Fig.~2, we show the results of the analysis in the ($\Omega_m,\sigma_8$) plane
for the cluster data alone.
Here, we marginalize over the other parameters, using a flat prior, and determine
the regions shown in the figure by finding where $\chi^2$ increases
by 1, 4, and 9, respectively, starting from the most likely set of values of the parameters.
As a consequence of this convention~\cite{PDB}, the projections of the allowed regions
onto each parameter give, respectively, the $1\sigma$, $2\sigma$, and $3\sigma$ intervals
for that parameter.
The filled contours are obtained taking into account only
statistical uncertainties (i.e., taking for $\mathcal{N}'_{{\rm obs},i}$ just the best
fit values listed in Table~III), while empty contours show the effect of including systematic
errors on the comoving numbers of clusters.
As it is apparent, the differences between the two cases are marginal.

Cluster data prefer large values of $\Omega_m$ and low values of $\sigma_8$
with respect to the standard $\Lambda$CDM concordance model. Indeed, we find the marginalized bounds
(including systematics),
\begin{eqnarray}
\label{sigma8limit}
\Omega_m \!\!& \geq &\! 0.38 \;\; (1\sigma \; \mbox{C.L.}) \ , \nonumber \\
\sigma_8 \!\!& \leq &\! 0.69 \;\; (1\sigma \; \mbox{C.L.}) \ ,
\end{eqnarray}
in slight tension at $1\sigma$ with those obtained from the 7-year WMAP
observations~\cite{WMAP7}: $\sigma_8 = 0.80 \pm 0.03 \; (68\% \, \mbox{C.L.})$ and
$\Omega_m = 0.265 \pm 0.011 \; (68\% \, \mbox{C.L.})$, although compatible at
$2\sigma$ confidence level. Note, however, that the WMAP's results are obtained
assuming a spatially flat
universe with cosmological constant and $H_0 = 71 \; \km/\s/\Mpc$.

Our results on $\Omega_m$ and $\sigma_8$ can be also compared to other recent cosmological results
coming from other galaxy cluster observations,
obtained using different strategies and cluster surveys.
In particular, studies of X-ray selected clusters, with masses exceeding a fixed
mass threshold and distributed over fixed redshift ranges, yield lower values of
$\Omega_m$ and larger values of $\sigma_8$, when compared with our limits~(\ref{sigma8limit}).
For example, Mantz et al.~\cite{Mantz2}
find $\Omega_m = 0.23 \pm 0.04$ and $\sigma_8 = 0.82 \pm 0.05$
in a model with a constant dark energy. 
Conversely, the analysis of cluster population
over low redshift ranges performed in~\cite{Schuecker}
gives a result compatible with ours, namely $\Omega_m = 0.34^{+0.09}_{-0.08}$ and
$\sigma_8 = 0.71^{+0.13}_{-0.16}$. Finally, the study of
the evolution of the $T_X$-based mass function~\cite{Vikhlinin}
within a model based on a cosmological constant,
gives $\Omega_m = 0.34 \pm 0.08$, in good agreement with our result.

From our fit we derive the $\sigma_8$ normalization
\begin{equation}
\label{sigma8Omegamrelation}
\sigma_8 \Omega_m^{\,1/3} = 0.49 \pm 0.06 \;\; (2\sigma \; \mbox{C.L.}) \ :
\end{equation}
it is worth comparing our results with those of Refs.\ \cite{Bahcall},
where the same set of cluster data was considered.
In Refs.\ \cite{Bahcall}, the authors analyze separately the data
of the first two bins and the data of the last two bins,
referring to low-redshift and high-redshift clusters, respectively (see Table~I).
They find $\sigma_8 \Omega_m^{0.6} = 0.33 \pm 0.03$
from a fit to the data of the first two bins, and
$\sigma_8 \Omega_m^{0.14} = 0.78 \pm 0.08$ from a fit to the data of the last two bins.
The above results in the ($\Omega_m,\sigma_8$) plane are shown
in the right panel of Fig.~2, superimposed to our result.
Requiring that both the above constraints be simultaneously satisfied,
the authors in Refs.\ \cite{Bahcall} find
$\Omega_m = 0.17 \pm 0.05 \; (1\sigma \; \mbox{C.L.})$ and
$\sigma_8 = 0.98 \pm 0.10 \; (1\sigma \; \mbox{C.L.})$
for the allowed 1$\sigma$ overlap region.


In conclusion, when our analysis is compared with that of Refs.\ \cite{Bahcall},
it is seen that the two analysis agree only marginally,
since our estimate favors relatively large values of $\Omega_m$ and low values
of $\sigma_8$. We trace the differences to our proper
treatment of cluster data, which depends on both redshift and cosmological parameters,
and to the correct calculation of the mass function which takes into account the
dependence of $\delta_c$ and $\Delta_v$ on redshift and cosmological parameters.
Note that in Refs.\ \cite{Bahcall} the Hubble parameter and the spectral index are fixed
(to the value $h=0.72$ and $n=1$) and $w=-1$. Also, the parameter $\kappa_\Delta$ in
Eq.~(\ref{MT1}) is assumed to be cosmology-independent and the value used is simply the arithmetic mean
of the values in Eq.~(\ref{kappanumbers}), namely $\kappa_\Delta = 0.98$.
The mass conversion is done by using the observed cluster profile
in the comoving radius range $R \in [0.5,2] \, h^{-1} \Mpc$.
Since some of the clusters we use in the analysis have comoving radii
exceeding the largest observed radius of $2 h^{-1} \Mpc$,
an extrapolation to higher value of $R$ is performed by assuming a
Navarro-Frenk-White profile for the virialized halo mass density (see Appendix~B).
Finally, the expressions for the critical density contrast, growth factor, virial
overdensity, and virial radius refer, in Refs.\ \cite{Bahcall}, to a matter dominated
universe with $\Omega_m = 1$.

Figure~3 shows the results of the analysis in the ($\Omega_m,\sigma_8$) plane
when we combine the data on galaxy cluster count with all the other cosmological data, discussed in Sec. III.
Closed thick contours show the allowed region obtained
combining cluster data (empty open thin contours) with all the remaining cosmological data
(filled vertical bands), the latter independent of $\sigma_8$.

There is a slightly tension between the values of $\Omega_m$ preferred by the clusters
and those preferred by the other cosmological probes, that are however compatible at 2$\sigma$ level.

For the sake of completeness, we quote the $1\sigma$ confidence limits for $(\Omega_m,\sigma_8)$, derived
from the joint analysis (clusters, BAO, CMB, Hubble parameter, and SNe):
\begin{eqnarray}
\label{w0wasigma8Omegamlimits}
\Omega_m \!\!& = &\! 0.28^{+0.03}_{-0.02} \;\;\;\;\; (1\sigma \; \mbox{C.L.}) \ , \\
\sigma_8 \!\!& = &\! 0.73^{+0.03}_{-0.03} \;\;\;\;\; (1\sigma \; \mbox{C.L.}) \ .
\end{eqnarray}

Let us conclude by giving the values of the minimum of the $\chi^2$ for clusters alone,
$\chi^2_{\rm CL,min} = 0.54$, for the remaining cosmological data,
$\chi^2_{\rm ALL-CL,min} = 539.2$, and for the joint analysis,
$\chi^2_{\rm ALL,min} = 542.5$. Taking into account the numbers of degrees of freedom,
of the same order of the values assumed by the $\chi^2$ or slightly larger,
these values confirm the goodness of our analysis.

\subsection{\normalsize{IVb. ($w_0$,$w_a$) results}}
\renewcommand{\thesection}{\arabic{section}}

Figures~4 shows the result of our analysis in the $(w_0,w_a)$ plane.
Empty and filled contours in the left and right panels refer to the same cases as
the left panel of Fig.~2 and Fig.~3, respectively.

It appears that current data on massive clusters do not allow one
to appreciably constrain the equation-of-state parameters $(w_0,w_a)$,
to either favor or rule out a cosmological
constant as dark energy (see left panel of Fig.~4).

The allowed $1\sigma$ confidence limits for $(w_0,w_a)$, derived
from the joint analysis (clusters, BAO, CMB, Hubble parameter, and SNe) are:
\begin{eqnarray}
\label{w0wasigma8Omegamlimits}
w_0 \!\!& = &\! -1.14^{+0.14}_{-0.16} \;\; (1\sigma \; \mbox{C.L.}) \ , \\
w_a \!\!& = &\! 0.85^{+0.42}_{-0.60} \;\;\;\;\; (1\sigma \; \mbox{C.L.}) \ . \\
\end{eqnarray}
The joint analysis is compatible (at 2$\sigma$ C.L.) with a cosmological constant
as dark energy.

Moreover, we find
\begin{equation}
\label{H0All}
H_0 = 69.1^{+1.3}_{-1.5} \;\, \km/\s/\Mpc \;\; (1\sigma \, \mbox{C.L.}) \ ,
\end{equation}
in agreement with a recent determination of the
Hubble constant from the Hubble Space Telescope
$H_0 = 73.8 \pm 2.4 \;\, \km/\s/\Mpc$~\cite{Riess},
and $\xi = -0.08^{+0.08}_{-0.90} \; (1\sigma \; \mbox{C.L.})$.

The results of the global analysis are practically independent of $n$
in the adopted range $[0.90,1.05]$.

For the sake of completeness, we report that from cluster data alone we
find that the results of the fit are almost independent of $n$, that they are only very weakly
dependent on $h$ in the adopted range $[0.6,0.8]$,
and that the pull is $\xi = -0.21^{+0.19}_{-0.74} \; (1\sigma \; \mbox{C.L.})$.

\subsection{\normalsize{IVc. XCDM results}}
\renewcommand{\thesection}{\arabic{section}}

Figure~5 shows the results of the analysis for the case $w_0 = w = \mbox{const}$
and $w_a = 0$ (the XCDM parametrization), in the $(\Omega_m,\sigma_8)$ and $(\Omega_m,w)$
planes.

{\it Only clusters.}-- As in the case of general evolving dark energy with
parameters $(w_0,w_a)$, cluster data prefer large values of $\Omega_m$,
and relatively small values of $\sigma_8$,
\begin{eqnarray}
\label{sigma8limitw}
\Omega_m \!\!& \geq &\! 0.43 \;\; (1\sigma \; \mbox{C.L.}) \ , \\
\sigma_8 \!\!& \leq &\! 0.67 \;\; (1\sigma \; \mbox{C.L.}) \ ,
\end{eqnarray}
and put very weak bounds on $w$ (see Fig.~5).

The left panel of Fig.~5 shows the correlation between $\Omega_m$ and $\sigma_8$ that can be
approximatively parameterized by
\begin{equation}
\label{sigma8Omegamrelationw}
\sigma_8 \Omega_m^{\,1/3} = 0.49 \pm 0.05 \;\; (2\sigma \; \mbox{C.L.}) \ ,
\end{equation}
with a slightly smaller error on $\sigma_8$ with respect to
Eq.~(\ref{sigma8Omegamrelation}).

{\it Combined data analysis.}-- The joint data analysis gives
\begin{eqnarray}
\label{w0wasigma8Omegamlimitsw}
\Omega_m \!\!& = &\! 0.28^{+0.03}_{-0.02} \;\;\;\;\; (1\sigma \; \mbox{C.L.}) \ , \\
\sigma_8 \!\!& = &\! 0.73^{+0.03}_{-0.03} \;\;\;\;\; (1\sigma \; \mbox{C.L.}) \ , \\
w \!\!& = &\! -0.96^{+0.08}_{-0.09} \;\; (1\sigma \; \mbox{C.L.}) \ .
\end{eqnarray}
Also, we find
\begin{equation}
\label{H0Allw}
H_0 = 69.0^{+1.4}_{-1.4} \;\, \km/\s/\Mpc \;\; (1\sigma \, \mbox{C.L.}) \ ,
\end{equation}
and $\xi = -0.08^{+0.08}_{-0.90} \; (1\sigma \; \mbox{C.L.})$. These results
are almost independent of $n$.

Let us quoting, also in this case, the values of the minimum of the $\chi^2$ for clusters alone,
$\chi^2_{\rm CL,min} = 0.56$, for the remaining cosmological data,
$\chi^2_{\rm ALL-CL,min} = 541.1$, and for the joint analysis,
$\chi^2_{\rm ALL,min} = 544.3$. As in the case of evolving dark energy,
these values confirm the goodness of our fits.

Observational constraints on the XCDM parametrization have been derived from many
different data sets, hence it provides a useful basis for comparing the discriminative
power of different data. It is well known that SNeIa apparent magnitude versus redshift,
BAO peak length scale, and CMB anisotropy data generally provide the most restrictive
constraints on cosmological parameters. Clearly, currently available massive cluster
evolution data is nowhere near as constraining as these data. However, cluster data results
in constraints that are comparable to those that follow from angular size versus redshift
data \cite{ADD} and lookback time data \cite{LT}, but are not as restrictive as those from
galaxy cluster gas mass fraction measurements \cite{GMF} or gamma-ray burst luminosity
observations \cite{GRB}. Over all, these constraints are approximately compatible
with each other and with the $\Lambda$CDM model, lending support to the belief that we
are converging on a standard cosmological model.


\section{\normalsize{V. Dark Energy Models}}
\renewcommand{\thesection}{\arabic{section}}

The above analysis shows that the constraints on the equation of state of dark energy
are only marginally affected by the inclusion of cluster data. Nevertheless, it is worth discussing
the impact of these constraints on different dark energy models as in~\cite{Barger}, since we
include more cosmological probes and upgraded data with respect to the analysis in~\cite{Barger}.

The advantage of the parametrization~(\ref{wz}) of the dark energy equation of state is twofold:
($i$) a number of dark energy models
can be adequately described by an equation of state of the form~(\ref{wz}) at recent
enough times (i.e., for $a$ near unity); and, ($ii$) at a given redshift (such that $a \sim 1$)
different classes of dark energy models correspond to different regions in the $(w_0,w_a)$
plane~\cite{Barger}. Indeed, roughly speaking, there exist four classes of dark energy models:
``thawing'' models, ``cooling'' models, ``barotropic'' fluids (all assumed to obey the null
energy condition $w \geq -1$), and ``phantom'' models (for which $w < -1$).\footnote{
This classification is not exhaustive since, as explained in~\cite{Barger},
some models, such as those with oscillating equations of state, do not fall into it.}
Introducing the quantity
\begin{equation}
\label{w'}
w' = \frac{dw}{d\ln a} \, , 
\end{equation}
the classification is as follows.

{\it Thawing models.}-- These satisfy the inequalities~\cite{Cadwell}
\begin{equation}
\label{Thawing}
1+w \lesssim w' \lesssim 3(1+w),
\end{equation}
and can arise in models of dark energy implemented by (cosmic) scalar fields,
such as axions or dilatons, which roll down towards the minimum of their potential.
Typical potentials are of the form $\phi^m$ ($m>0$), with $\phi$ being the scalar field.
The bounds~(\ref{Thawing}) are valid for $(1+w) \ll 1$ so, following
Ref.~\cite{Cadwell}, we assume $w \lesssim -0.8$ as a practical limit
of applicability.
It should be noticed, as recently pointed out in Ref.~\cite{Gupta}, that is some scalar field models, the 
thawing model parameter space is slightly larger than the one
defined by Eq.~(\ref{Thawing}).
 
{\it Cooling models.}-- As for the case of thawing models, cooling models may arise
in scalar field models of dark energy. Typical scalar potentials are of the form
$\phi^{-m}$ ($m>0$). They lie in the region~\cite{Cadwell,Sherrer}
\begin{equation}
\label{Cooling}
-3(1-w)(1+w) < w' \lesssim \epsilon(z) \, w(1+w) \ ,
\end{equation}
where $\epsilon$ is a function of the redshift and an investigation of a variety
of scalar-field cooling models indicates that $\epsilon(1) \simeq 0.2$~\cite{Cadwell}.
The upper bound in Eq.~(\ref{Cooling}) is valid for $(1+w) \ll 1$ ~\cite{Cadwell} so,
for this bound, we assume $w \lesssim -0.8$.
Cooling models may arise in models of dynamical supersymmetry breaking and supergravity.
So-called $k$-essence models~\cite{Mukhanov} with a nonlinear kinetic term belong to this class~\cite{Chiba}.


{\it Barotropic fluids.}-- These are fluids whose pressure $p$ depends only on
the energy density $\rho$. Assuming that $c_s^2 = dp/d\rho > 0$,
barotropic fluids satisfy the inequality~\cite{Sherrer}
\begin{equation}
\label{Barotropic}
w' < 3w(1+w) \ ,
\end{equation}
and include (original~\cite{Moschella} and generalized~\cite{Bertolami}) Chaplygin gas models.

{\it Phantom models.}-- These are models which do not obey the null energy condition
(see, however, Ref.~\cite{Barger}), and fall into the region
\begin{equation}
\label{Phantom}
w < -1 \ .
\end{equation}


\begin{figure*}[t]
\begin{center}
\hspace*{-1.9cm}
\includegraphics[clip,width=21.3cm,bb=1 10 570 340]{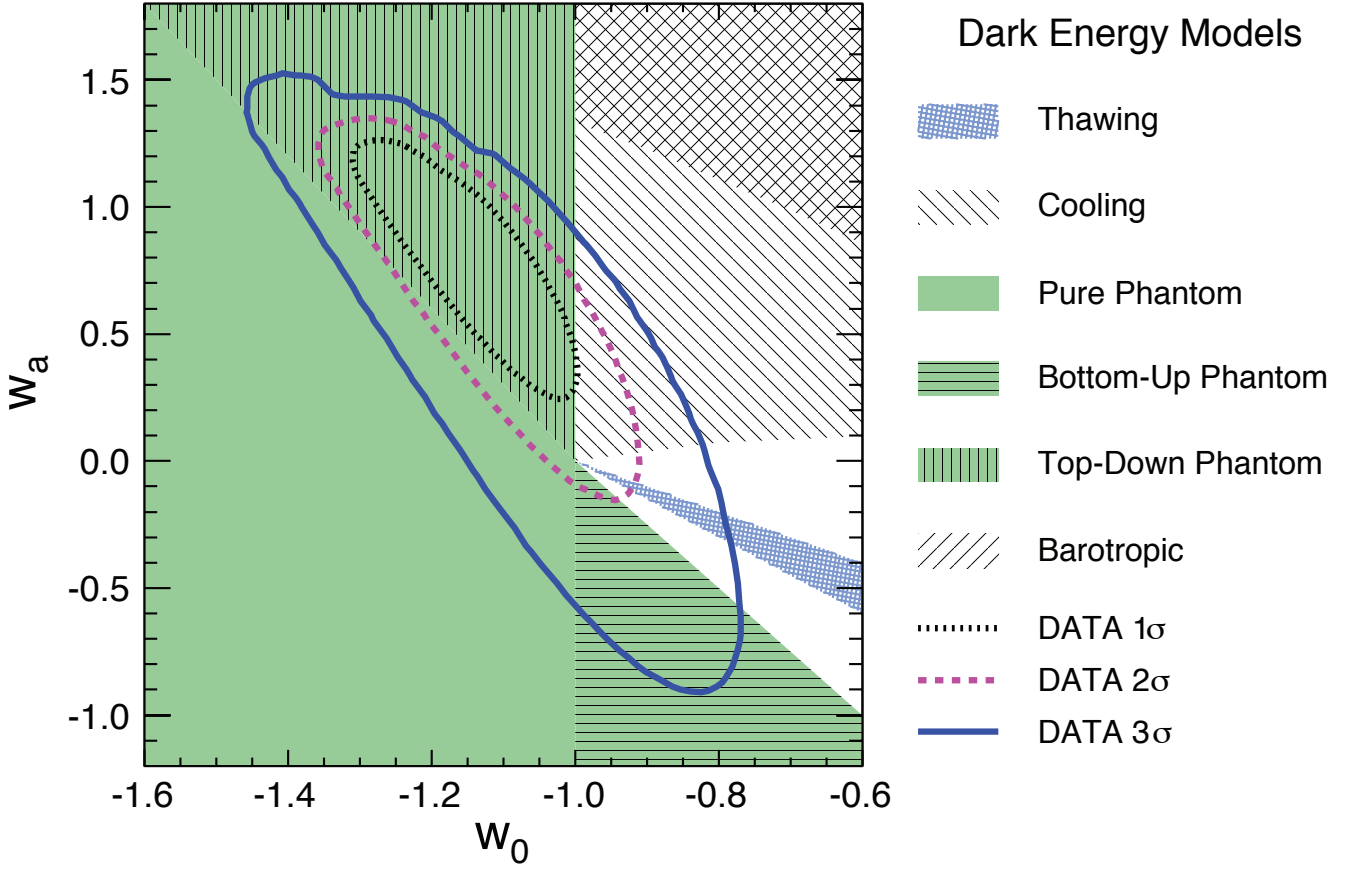}
\vspace*{-0.8cm}
\caption{$1\sigma$, $2\sigma$, and $3\sigma$ confidence
level contours in the $(w_0,w_a)$ plane from the joint analysis of galaxy cluster number count
(with systematics), BAO, CMB, Hubble parameter, and SNe (the same as in the right panel of Fig.~4).
The shaded areas represent different types of evolving dark energy models
according to the classification of Ref.\ \cite{Barger}.}
\end{center}
\end{figure*}


To each of the above models, we can associate a specific region in the $(w_0,w_a)$ plane
at a given reference time. Following Ref.~\cite{Barger}, we take as reference time that
corresponding to $z=1$ which, roughly speaking, is when dark energy is expected to start
to dominate over nonrelativistic matter.
Phantom models at $z=1$ can be split in two classes:
``Pure phantom'' models which did not cross the phantom divide line $w=-1$ recently,
\begin{equation}
\label{Phantom1}
w_0 < -1, \;\;\;\; w(z=1) < -1 \ ,
\end{equation}
and models that crossed $w=-1$ from a lower value to a higher value,
\begin{equation}
\label{Phantom2}
w_0 > -1, \;\;\;\; w(z=1) < -1 \ ,
\end{equation}
which we dub ``bottom-up phantom'' models.

Finally, we also consider
models that crossed $w=-1$ from a higher value to a lower value:
\begin{equation}
\label{Phantom3}
w_0 < -1, \;\;\;\; w(z=1) > -1 \ .
\end{equation}
These models, which we dub ``top-down phantom'' models,
are phantom today ($z=0$) and non-phantom at $z=1$.

The $(w_0,w_a)$ plane containing all the above regions is presented in Fig.\ 6,
together with the regions allowed by data and discussed in Section V.
\footnote{There are a few disadvantages of the parametrization of Eq.\ (\ref{wz}): ($i$) This
two parameter ($w_0, w_a$) parametrization has one more parameter than the simplest
consistent and physically motivated scalar field dark energy model, \cite{PR88, RP88},
thus making it more difficult to constrain model parameters with observational
data; and, ($ii$) even so, the parametrization is not physically complete as additional
information must be provided if one is interested in the evolution of energy density
and other spatial inhomogeneities, \cite{XCDMincompleteness}.}

Figure~6 shows the $1\sigma$, $2\sigma$, and $3\sigma$ confidence level contours
in the $(w_0,w_a)$ plane (obtained from the global
analysis) superimposed on the regions representing different
types of evolving dark energy models according to the classification
of Ref.\ \cite{Barger}. At the $1\sigma$ level,
phantom models of evolving dark energy of top-down type are slightly
favored over cooling models and considerably preferred over thawing, pure,
and bottom-up phantom models. Non-phantom barotropic fluids are ruled out.


\section{\normalsize{VI. Conclusions}}
\renewcommand{\thesection}{\arabic{section}}

We have constrained cosmological parameters by using X-ray
temperature data of massive galaxy clusters in the redshift range $0.05 \lesssim z \lesssim 0.83$
with masses within a comoving radius of $1.5 h^{-1}\Mpc$ greater than the
fiducial value $8 \times 10^{14} h^{-1} M_\odot$.

In this analysis, we have accounted for the dependence of quantities related to cluster physics
--- such as the critical density contrast, the growth factor, the mass conversion factor, the virial
overdensity, and the virial radius --- on the cosmological model parameters $\Omega_m,
\sigma_8,n,w_0,w_a$, and $H_0$. We have also taken into account the dependence on redshift and
cosmological parameters of the mass-temperature relation which allow us to convert
the observed cluster X-ray temperatures into cluster masses and to calculate the cluster number counts.

The analyses show that cluster data prefer small values of the amplitude of mass
fluctuations $\sigma_8$,
\begin{equation}
\label{C1}
\mbox{\it only clusters:} \;\;\; \sigma_8 \leq 0.69 \;\; (1\sigma \, \mbox{C.L.}) \ ,
\end{equation}
as well as large values of nonrelativistic matter energy density,
\begin{equation}
\label{C2}
\mbox{\it only clusters:} \;\;\; \Omega_m \geq 0.38 \;\; (1\sigma \, \mbox{C.L.}) \ .
\end{equation}
The above bounds are in slight tension at $1\sigma$ with those obtained from the 7-year WMAP
observations, although compatible at $2\sigma$ confidence level. In addition, we have found
the following normalization of
$\sigma_8$:
\begin{equation}
\label{C3}
\mbox{\it only clusters:} \;\;\; \sigma_8 \Omega_m^{\,1/3} = 0.49 \pm 0.06 \;\; (2\sigma \, \mbox{C.L.}) \ .
\end{equation}
We have found that currently available cluster data do not tightly constrain the
dark energy equation of state, and that a cosmological constant is consistent with these
observations.

Cluster data alone are not sensitive to the value of the index $n$ of the power-law
power spectrum of the density perturbations, and are only very weakly dependent
on the Hubble constant $H_0$.

In order to break the $\Omega_m$-$\sigma_8$ degeneracy and put more stringent constraints
on cosmological parameters, we have combined cluster data with
BAO peak length scale observations, CMB anisotropy data, Hubble constant and Hubble parameter
measurements, and type Ia supernova magnitude-redshift observations. In this case we find
\begin{equation}
\label{C4}
\mbox{\it all data:} \;\;\; \sigma_8 = 0.73^{+0.03}_{-0.03} \;\; (1\sigma \, \mbox{C.L.})
\end{equation}
and
\begin{equation}
\label{C5}
\mbox{\it all data:} \;\;\; \Omega_m = 0.28^{+0.03}_{-0.02} \;\; (1\sigma \, \mbox{C.L.}) \ ,
\end{equation}
which are in good agreement with previous constraints in the literature
(such as those coming from WMAP).

Regarding the equation-of-state parameters of dark energy, we find
\begin{equation}
\label{C6}
\mbox{\it all data:} \;\;\; w_0 = -1.14^{+0.14}_{-0.16} \ ,
                      \;\;  w_a = 0.85^{+0.42}_{-0.60} \;\; (1\sigma \; \mbox{C.L.}) \ ,
\end{equation}
which indicates that the joint analysis is consistent with a cosmological constant.
Moreover, the combination of all data is almost insensitive to $n$,
and constrains the Hubble parameter to the range,
\begin{equation}
\label{C7}
\mbox{\it all data:} \;\;\; H_0 = 69.1^{+1.3}_{-1.5} \;\, \km/\s/\Mpc \;\; (1\sigma \, \mbox{C.L.}) \ ,
\end{equation}
consistent with recent bounds from Hubble Space Telescope observations.

Similar results are found in the case of constant equation-of-state parameter
time-varying dark energy (the XCDM parametrization).

Our results suggest that, among models of dark energy with varying equation of state
(i.e., with $w_a \neq 0$),
the top-down phantom models, for which the equation of state crossed the phantom
divide line from a higher value to a lower value, are preferred over non-phantom thawing and
cooling models. Finally, non-phantom barotropic fluids are excluded as models of dark energy.

While currently available massive cluster data do not constrain cosmological parameters
as tightly as do SNeIa apparent magnitude versus redshift measurements, or CMB anisotropy
data, or BAO peak length scale observations, the cluster measurements do provide constraints
comparable to those from some of the other available data sets. They also play a useful
role in constraining cosmological parameters when used in conjunction with other data. More
importantly, we look forward to superior quality near-future massive cluster data, and
anticipate the significantly more restrictive parameter constraints that will result
from using the techniques we have developed here.


\vspace*{0.3cm}

\begin{acknowledgments}
We would like to thank A. Natarajan for useful comments.
T.K.\ acknowledges partial support from Georgian National Science Foundation grant ST08/4-422,
Swiss National Science Foundation SCOPES grant 128040, NASA grant NNXlOAC85G, NSF grant AST-1109180,
the ICTP associate membership program, and the Universit\`{a} Degli Studi Di Bari for hospitality.
B.R.\ acknowledges support from DOE grant DEFG03-99EP41093 and NSF grant AST-1109275.
\end{acknowledgments}


\section{\normalsize{Appendix A. Critical Density Contrast and Growth Factor}}


\begin{figure*}[t]
\vspace*{-2cm}
\begin{center}
\includegraphics[clip,width=1.0\textwidth]{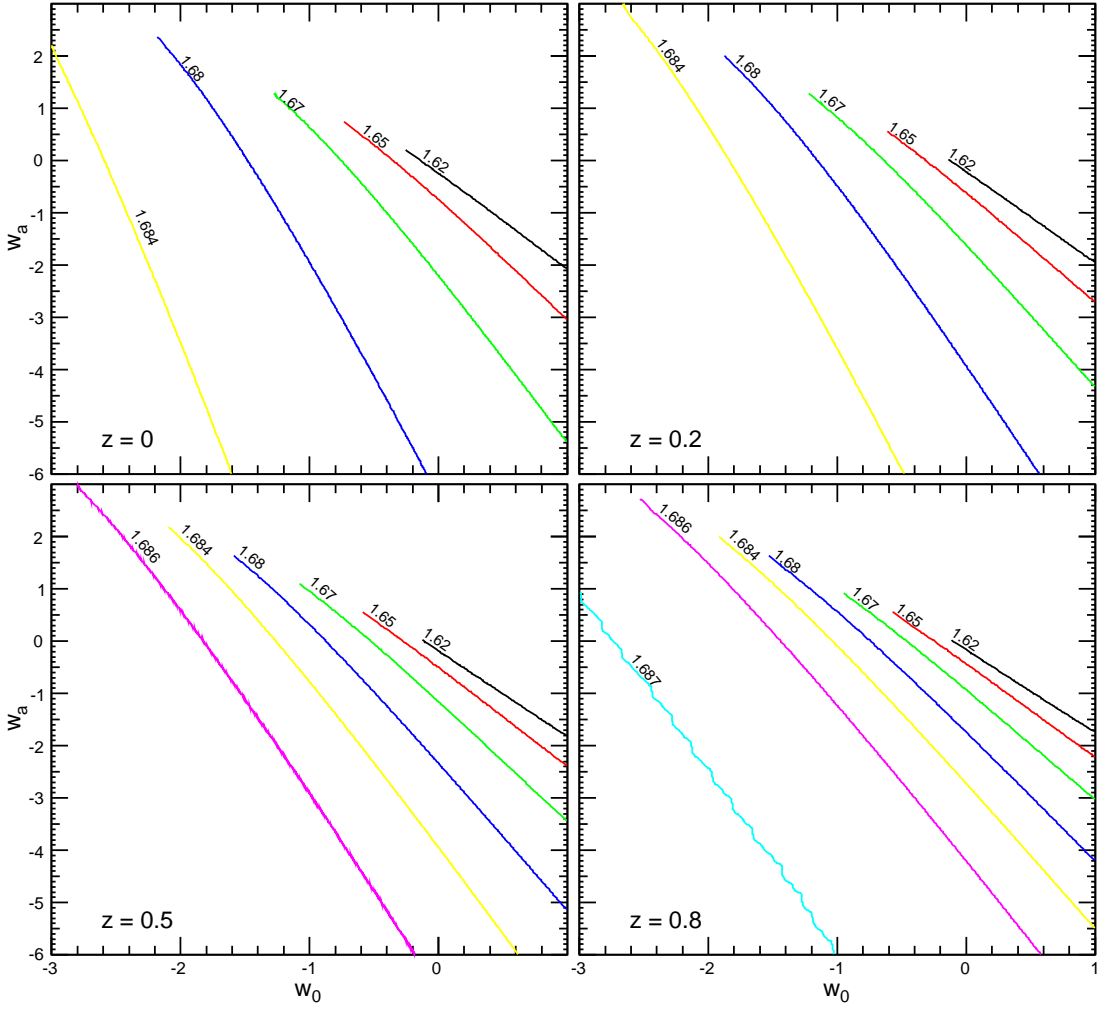}
\vspace*{-2cm}
\caption{$\delta_c$-isocontours in the $(w_0,w_a)$-plane for different values of
the redshift and for $\Omega_m = 0.27$.}
\end{center}
\end{figure*}


{\it Critical Density Contrast.}-- The critical density contrast $\delta_c$ depends
on the redshift and on the cosmology and can be evaluated, using the approach of
Ref.\ \cite{Pace} (see also Ref.\ \cite{Mota}), as follows. Consider the full nonlinear
equation describing the evolution of the density contrast:
\begin{equation}
\label{deltaNL}
\delta'' + \left( \frac{3}{a} + \frac{E'}{E} \right) \! \delta' - \frac43 \frac{\delta'^2}{1+\delta}
- \frac32 \frac{\Omega_m}{a^5 E^2} \, \delta(1+\delta) = 0 \ ,
\end{equation}
where a prime indicates differentiation with respect to the scale factor $a$ and
\begin{equation}
\label{deltaDEF}
\delta = \frac{\delta \rho_m}{\rho_m} = \frac{\rho_{\rm cluster} - \rho_m}{\rho_m}
\end{equation}
is the density contrast with $\rho_{\rm cluster}$ being the cluster matter density.
Since the above equation describes the nonlinear growth of the
density contrast, its value at some chosen collapse time $t_{\rm collapse}$ diverges.
The critical density contrast $\delta_c$ at the time $t_{\rm collapse}$ is, by definition,
the value of the density contrast at the time $t_{\rm collapse}$ obtained by solving the linearized
version of Eq.~(\ref{deltaNL}), namely
\begin{equation}
\label{delta}
\delta'' + \left( \frac{3}{a} + \frac{E'}{E} \right) \! \delta' - \frac32 \frac{\Omega_m}{a^5 E^2} \, \delta = 0 \ ,
\end{equation}
with boundary conditions for $\delta$ such that, the same boundary conditions
applied to the nonlinear equation~(\ref{deltaNL}) makes $\delta$ divergent at $t_{\rm collapse}$.
Following Ref.~\cite{Pace}, we take for the initial derivative of $\delta$, $\delta'(a_i)$,
the value $\delta'(a_i) = 5 \times 10^{-5}$ where $a_i \equiv 5 \times 10^{-5}$,
while the initial value of the density contrast, $\delta(a_i)$, is found by searching for the
value of $\delta(a_i)$ such that $\delta$ diverges at the time $t_{\rm collapse}$. We assume
(as in Ref.~\cite{Pace}) that the divergency is achieved, numerically, when $\delta$ exceeds
the value $10^7$.

In Fig.~1, we plot the critical density contrast, $\delta_c$, as a function of the redshift
for different values of $(w_0,w_a)$. For large redshifts --- where the effects of dark energy
become negligible compared to those of nonrelativistic matter --- the Universe effectively
approaches the Einstein-de Sitter model where the critical density contrast is independent of
the redshift and is $\delta_c = (3/20)(12\pi)^{2/3} \simeq 1.686$~\cite{Weinberg}.
In Fig.~7, we show the $\delta_c$-isocontours in the $(w_0,w_a)$-plane for different values of
the redshift and for $\Omega_m = 0.27$.

{\it Growth Factor.}-- The $z$-dependent part of the matter power spectrum ---
the growth factor $D(z)$ --- is
\begin{equation}
\label{D}
D(z) = \frac{\delta(z)}{\delta(0)} \ ,
\end{equation}
and satisfies the linearized equation~(\ref{delta}). The boundary conditions we impose are
$D(a=1) = 1$ and $D(a=a_i) = a_i$, where as before $a_i \equiv 5 \times 10^{-5}$. In Fig.~1
we plot the growth factor, $D(z)$, as a function of the redshift for different values of $(w_0,w_a)$.


\section{\normalsize{Appendix B. Mass Conversion}}

The virial mass in the PS or ST parametrization needs to be expressed as a function
of the observed mass $M'$ within a reference comoving radius of $R_0' = 1.5 h^{-1}\Mpc$.
As described in Section IIIb, what we really need is the fiducial virial mass $M_0$
as a function of the fiducial mass within $R_0'$ adopted in the observation,
namely $M_0' = 8 \times 10^{14} h^{-1} M_\odot$. We use the following procedure
to accomplish this. We first determine the physical virial radius $r_{v,0}$,
within which the virial mass $M_0$ is contained, through the relation
\begin{equation}
\label{Mvirial}
M_0(z) = \frac{4\pi}{3} \, r_{v,0}^3(z) \rho_m(z) \Delta_v(z) =
\frac{4\pi}{3} \, R_{v,0}^3 \rho_m^{(0)} \Delta_v(z) \ ,
\end{equation}
where $R_{v,0} = (1+z) \, r_{v,0}$, and
$\Delta_v$ is the virial overdensity and is discussed in Appendix C.
We then scale the virial mass $M_0$ to the $1.5 h^{-1}\Mpc$ comoving radius
assuming a Navarro-Frenk-White profile for the virialized halo mass density~\cite{NFW}:
\begin{equation}
\label{NFW}
\rho_{\rm cluster}(r) = \frac{4\rho_{\rm cluster}(r_s)}{(r/r_s)(1+r/r_s)^2} \ ,
\end{equation}
where $r$ is the physical radial distance and $r_s$ is a physical scale radius.

Technically the procedure is as follows.
From Eq.~(\ref{NFW}) we can obtain the mass $M(<r)$ contained in the physical radius $r$:
\begin{eqnarray}
\label{Mr}
M(<r) \!\!& = &\!\! 4\pi \int_0^r dr' r'^2 \rho_{\rm cluster}(r') \nonumber \\
\!\!& = &\!\! 16 \pi \rho_{\rm cluster}(r_s) \, r^3 f(r_s/r) \ ,
\end{eqnarray}
where
\begin{equation}
\label{f}
f(x) = x^3 \! \left[ \ln \left( \frac{1+x}{x} \right) - \frac{1}{1+x} \right] \ .
\end{equation}
Then applying Eq.~(\ref{Mr}) to the mass $M_0$ within the physical virial radius $r_{v,0}$,
and to the mass $M_0'$ within the physical radius $r'_0 = R_0'/(1+z)$ corresponding to the
comoving radius $R'_0$, and taking the ratio of these two equations, we get
\begin{eqnarray}
\label{M/M}
\frac{M_0}{M_0'} = \left( \frac{R_{v,0}}{R'_0} \right)^{\!3}
\frac{f\!\left(1/c_{v,0} \right)}
{f\!\left(R_{v,0}/ c_{v,0}  R'_0 \right)} \ .
\end{eqnarray}
Here
\begin{equation}
\label{c0}
c_{v,0} = \frac{r_{v,0}}{r_s}
\end{equation}
is the ``concentration parameter''
\begin{equation}
\label{c}
c_v = \frac{r_v}{r_s} \
\end{equation}
--- the ratio between the physical virial radius $r_v$ and the physical scale radius $r_s$ ---
evaluated at the physical virial radius $r_{v,0}$.

Both $R_{v,0}$ and $c_{v,0}$ in Eq.~(\ref{M/M}) are functions of $M_0$.
From Eq.~(\ref{Mvirial}) we can express $R_{v,0}$ as a function of $M_0$.
For $c_{v,0}$, we proceed as follows. First we consider the expression
for the concentration parameter found by Bullock et al.~\cite{Bullock}
in their $N$-body simulation:
\begin{equation}
\label{cBullock}
c_v(M_v) = \frac{B}{1+z} \, \left[ \frac{M_v(0)}{M_*(0)} \right]^{-\beta} \ .
\end{equation}
Here $B \simeq 9$, $\beta \simeq 0.13$,
\footnote{To be precise, $B$ and $\beta$ can (weakly) depend on the cosmology and the values used here
are those found in Ref.~\cite{Bullock} where a cosmology with cosmological constant was assumed.}
%
$M_v$ is the mass within a physical virial radius $r_v$,
\begin{equation}
\label{Mv}
M_v(z) = \frac{4\pi}{3} \, r_v^3(z) \rho_m(z) \Delta_v(z) \ ,
\end{equation}
and $M_*(z)$ is a fiducial mass defined by
\begin{equation}
\label{Mstarz}
\sigma(R_*(z),z) = \delta_c(z) \ ,
\end{equation}
where the comoving radius $R_*(z)$ is defined by
\begin{equation}
\label{Rstarz}
M_*(z) = \frac{4\pi}{3} \, R_*^3(z) \, \rho_m^{(0)} \ .
\end{equation}
Then, evaluating Eq.~(\ref{cBullock}) for $M_v = M_0$
[which corresponds to evaluating Eq.~(\ref{c}) for $r_v = r_{v,0}$], we have
\begin{equation}
\label{c0Bullock}
c_{v,0} = c_v(M_0) = \frac{c_{v,0}'}{\gamma} \left( \frac{M_0}{M_0'} \right)^{\!-\beta} \ ,
\end{equation}
where
\begin{eqnarray}
\label{c'0}
c_{v,0}' = c_v(M_0') = \frac{B}{1+z} \, \left[ \frac{M_0'}{M_*(0)} \right]^{-\beta}
\end{eqnarray}
is, formally, the concentration parameter (\ref{cBullock}) evaluated at
$M_0'$ [see Eq.~(\ref{cBullock})], and
\begin{eqnarray}
\label{gamma}
\gamma(z) = \left[ \frac{\Delta_v(z)}{\Delta_v(0)} \right]^{-\beta} \ .
\end{eqnarray}
Taking into account Eqs.~(\ref{Sigma}), (\ref{sigmaSigma}), and (\ref{Mstarz}), we find that the quantity
$M_*(0)$ does not depend on $M_0$ and is defined by
\begin{equation}
\label{equality}
\Sigma(R_*(0)/R_0) = \frac{\delta_c(0)}{\sigma_8} \ .
\end{equation}
Finally, inserting Eqs.~(\ref{Mvirial}) and (\ref{c0Bullock}) in Eq.~(\ref{M/M}),
we obtain the equation that gives $M_0$ as a function of $M_0'$ [namely the equation
defining the function $g$ in Eq.~(\ref{vM})]:
\begin{equation}
\label{M0/M0}
f \! \left( R_{1.5} \, \frac{\gamma}{c_{v,0}'} \left( \frac{M_0}{M_0'} \right)^{\!\beta + 1/3} \right) =
R_{1.5}^3 \, f \! \left( \frac{\gamma}{c_{v,0}'} \left( \frac{M_0}{M_0'}\right)^{\!\beta} \right) \ ,
\end{equation}
where
\begin{equation}
\label{R15}
R_{1.5}(z) = \frac{\left[ 3M_0'/4\pi \rho_m^{(0)} \Delta_v(z) \right]^{1/3}}{1.5 h^{-1}\Mpc}
\end{equation}
is the comoving virial radius corresponding to the mass $M_0'$, normalized to $1.5 h^{-1}\Mpc$.

In a flat universe dominated by nonrelativistic matter (the Einstein-de Sitter model), the comoving
virial radius corresponding to the mass $M_0' = 8 \times 10^{14} h^{-1} M_\odot$ is about
$1.57 h^{-1}\Mpc$, which gives $R_{1.5} \simeq 1$. This in turns means, using Eq.~(\ref{M0/M0}),
that $M_0/M_0' \simeq 1$, namely the virial mass contained in a sphere of comoving virial radius
of $1.5 h^{-1}\Mpc$ is approximatively $M_0'$.
Due to the presence of the third root in Eq.~(\ref{R15}), the function
$R_{1.5}(z)$ is fairly insensitive to changes in the adopted cosmology
[through $\rho_m^{(0)}$ and $\Delta_v(z)$]. Therefore, also in the case of a generic cosmology
with evolving dark energy, we expect values of $M_0$ near to $M_0'$.

In Fig.~1, we plot the ratio $M_0/M_0'$ as a function of the redshift
for different values of $(w_0,w_a)$. $M_0$, as argued, turns to be of order
(of a few times) $M_0'$.


\section{\normalsize{Appendix C. Virial Overdensity and Virial Radius}}

{\it Virial Overdensity.}-- The virial overdensity is the ratio of the cluster
mass density and the background matter density
at the time of virialization:
\begin{equation}
\label{Deltav}
\Delta_v(z=z_v) = \frac{\rho_{\rm cluster}(z_v)}{\rho_m(z_v)} \ ,
\end{equation}
where $z_v$ is the redshift at the time of virialization.

Using the fact that the cluster mass density is
\begin{equation}
\label{Mcluster}
\rho_{\rm cluster}(z) = \frac{3M_{\rm cluster}}{4\pi r^3(z)} \ ,
\end{equation}
where $M_{\rm cluster}$ is the halo cluster mass and $r(z)$ the physical halo radius,
we can rewrite Eq.~(\ref{Deltav}) as
\begin{equation}
\label{Deltav2}
\Delta_v(z=z_v) = \left( \frac{r_{\rm ta}}{r_v} \right)^{\!3}
\left( \frac{1 + z_{\rm ta}}{1+ z_v} \right)^{\!3} \zeta \ ,
\end{equation}
where $r_{\rm ta}$ and $r_v$ are, respectively,
the physical radii of the halo cluster at turn-around and at virialization
($r_v$ is, in other words, the physical virial radius of the halo cluster),
while $z_{\rm ta}$ and
\begin{equation}
\label{zeta}
\zeta = \frac{\rho_{\rm cluster}(z_{\rm ta})}{\rho_m(z_{\rm ta})}
\end{equation}
are, respectively, the redshift and virial overdensity at the time of turn around.


\begin{figure*}[t]
\vspace*{-2cm}
\begin{center}
\includegraphics[clip,width=1.0\textwidth]{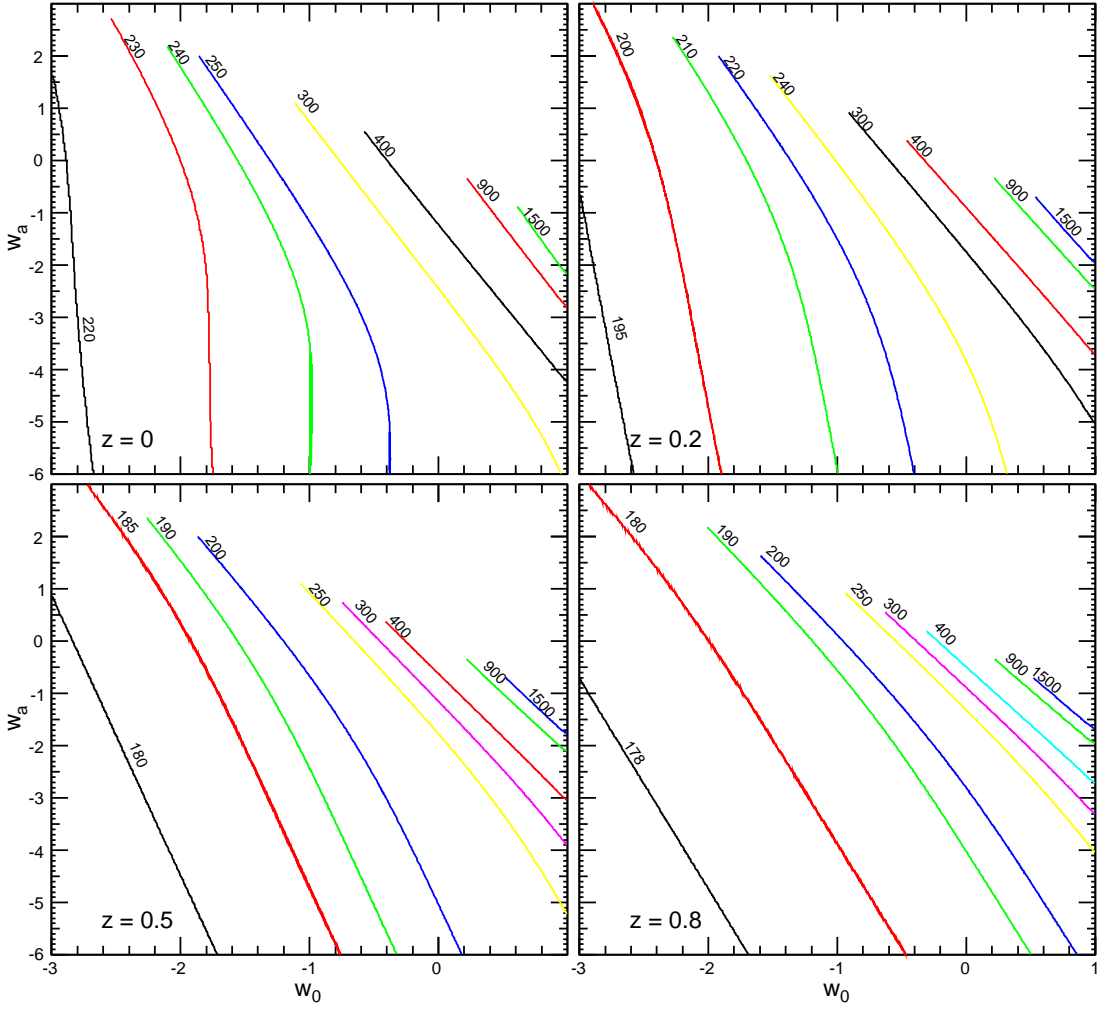}
\vspace*{-2cm}
\caption{$\Delta_v$-isocontours in the $(w_0,w_a)$-plane for different values of
the redshift. We fixed $\Omega_m = 0.27$.}
\end{center}
\end{figure*}


The redshift at the virialization time is
\begin{equation}
\label{zcollapse}
z_v = z_{\rm collapse},
\end{equation}
and follows from the standard assumption that clusters virialize at the time of collapse.

To find the redshift at the time of turn-around, $z_{\rm ta}$, and to obtain the virial
overdensity at the turn-around time, $\zeta$, we follow the procedure of Ref.\ \cite{Pace}.
First, they observed that the quantity $(\delta + 1)/a^3$, where $\delta$ is the density
contrast that satisfies the nonlinear Eq.~(\ref{deltaNL}), is proportional to $1/r^3(z)$:
\begin{equation}
\label{delta+1}
\frac{\delta +1}{a^3} = \frac{3M_{\rm cluster}}{4\pi \rho_m^{(0)}} \, \frac{1}{r^3(z)} \ ,
\end{equation}
where $r(z)$ is the collapsing sphere's radius.
[It is straightforward to get the above equation by using Eqs.~(\ref{deltaDEF}) and (\ref{Mcluster}).]
Then since $r(z)$ assumes the maximum value at the turn-around time,
the time of turn-around is found by minimizing the quantity $(\delta + 1)/a^3$,
where $\delta$ is the solution of Eq.~(\ref{deltaNL}) obtained by imposing the
boundary conditions used in Appendix A.

Once the turn-around time is found, the virial overdensity at the turn-around time
is simply given, from Eqs.~(\ref{deltaDEF}) and (\ref{zeta}), by
\begin{equation}
\label{zsol}
\zeta = \delta(z_{\rm ta}) + 1 \ ,
\end{equation}
where $\delta$ is the solution of Eq.~(\ref{deltaNL}).

The ratio of the virial radius to the turn-around radius, $r_v/r_{\rm ta}$,
as a function of cosmological parameters is analyzed below.

In Fig.~1 we plot the virial overdensity at the time of virialization, $\Delta_v$,
as a function of the redshift for different values of $(w_0,w_a)$. In an
Einstein-de Sitter model (i.e., for $\Omega_m = 1$), the standard assumption that
$t(z_v) \simeq 2t(z_{\rm ta})$ together with the fact that $r_v = r_{\rm ta}/2$ and
$\zeta = (3\pi/4)^2$, gives $\Delta_v \simeq 18\pi^2 \simeq 177.653$ (independent of the redshift).
Indeed, each curve in Fig.~1 approaches this value for a sufficiently large value of the
redshift since, as already noted, the Universe then enters the Einstein-de Sitter regime
where the effects of dark energy become subdominant with respect to those of nonrelativistic
matter.

In Fig.~8, we show the $\Delta_v$-isocontours in the $(w_0,w_a)$-plane for different values of
the redshift and for $\Omega_m = 0.27$.

{\it Virial Radius.}-- In order to ``estimate'' the quantity $r_v/r_{\rm ta}$,
we apply energy conservation and the virial theorem to the spherical collapse of
the cluster halo.

We start by considering the total gravitational potential energy $U(r)$ of a
sphere of radius $r$ containing the cluster mass $M_{\rm cluster}$ and dark energy:
\begin{equation}
\label{U}
U(r,z) = U_{mm}(r) + U_{m \rm DE}(r,z) \ ,
\end{equation}
where
\begin{equation}
\label{Umm}
U_{mm}(r) = -\frac{3}{5} \, \frac{GM_{\rm cluster}^2}{r}
\end{equation}
is the familiar gravitational potential self-energy of a sphere of nonrelativistic matter,
and
\begin{equation}
\label{UmDE}
U_{m \rm DE}(r,z) = -\frac{4\pi}{5} \, GM_{\rm cluster} \, [1+3w(z)] \, \rho_{\rm DE}(z) \, r^2
\end{equation}
is the gravitational potential energy of interaction between nonrelativistic matter and
dark energy.\footnote{
Here, we are considering the case where the dark energy does not cluster and does not virialize,
so that the only terms in the potential energy which are relevant for energy conservation and
virialization are those in Eq.~(\ref{U})~\cite{Wang}.}

Since the potential energy $U_{m \rm DE}(r,z)$ depends explicitly on the time, the system
under consideration is not conservative. Therefore, neither energy conservation nor the virial
theorem can be applied.\footnote{
It is straightforward to show that the only case where $U_{m \rm DE}(r,z)$ is $z$-independent,
and the system is conservative, is that of the cosmological constant, namely $w(z) = -1$ for all
times.}

In order to get a conservative system, Wang~\cite{Wang} has suggested replacing
the $z$-dependent quantity $[1+3w(z)] \, \rho_{\rm DE}(z)$ with the same quantity
evaluated at the turn-around time.

Here we propose defining an effective potential energy, which does not depend explicitly
on the time, as
\begin{equation}
\label{Ueff}
U^{\rm (eff)}(r) = U_{mm}(r) + U^{\rm (eff)}_{m \rm DE}(r) \ ,
\end{equation}
where
\begin{equation}
\label{UmDEeff}
U^{\rm (eff)}_{m \rm DE}(r) =
-\frac{4\pi}{5} \, GM_{\rm cluster} \, \langle \, [1+3w(z)] \, \rho_{\rm DE}(z) \, \rangle \, r^2  \\
\end{equation}
is the effective potential energy of interaction and $\langle ... \rangle$ is an operator
that when applied to a $z$-dependent function $\psi(z)$ gives a $z$-independent quantity, i.e.
\begin{equation}
\label{g}
\frac{d \langle \, \psi(z) \, \rangle}{dz} = 0 \ .
\end{equation}
The action of the $\langle ... \rangle$-operator is specified below.

The introduction of the effective energy potentials (\ref{Ueff}) and (\ref{UmDEeff}),
allow us to use the energy conservation theorem that, applied at the times of
virialization and turn-around, gives
\begin{equation}
\label{Energy}
\mathcal{K}(r_v) + U^{\rm (eff)}(r_v) = U^{\rm (eff)}(r_{\rm ta}) \ ,
\end{equation}
where $\mathcal{K}(r_v)$ is the kinetic energy at the virialization time.

Using the virial theorem
%
%
\begin{equation}
\label{Virial}
\mathcal{K}(r_v) = \left( \frac{r}{2} \, \frac{dU^{\rm (eff)}(r)}{dr} \right)_{\!r = r_v} \ ,
\end{equation}
the energy conservation equation (\ref{Energy}) takes the form
\begin{equation}
\label{U=U}
\frac{1}{2} \, U_{mm}(r_v) + 2 U^{\rm (eff)}_{m \rm DE}(r_v)  =
U_{mm}(r_{\rm ta}) + U^{\rm (eff)}_{m \rm DE}(r_{\rm ta}) \ .
\end{equation}
Taking into account Eqs.~(\ref{Mcluster}) and (\ref{zeta}), the mass of the cluster is
\begin{equation}
\label{Mzeta}
M_{\rm cluster} = \frac{4\pi}{3} \, r_{\rm ta}^3 \, \rho_m(z_{\rm ta})  \, \zeta \ ,
\end{equation}
so that Eq.~(\ref{U=U}) reads
\begin{equation}
\label{x}
4 \eta q(1+3w_{\rm ta}) \! \left( \frac{r_v}{r_{\rm ta}} \right)^{\!3} -
2 \left[1 + \eta q(1+3w_{\rm ta}) \right] \! \left( \frac{r_v}{r_{\rm ta}} \right) + 1 = 0 \ .
\end{equation}
Here $w_{\rm ta} = w(z_{\rm ta})$ and we have defined, following Ref.~\cite{Wang}, the quantity
\begin{eqnarray}
\label{q}
q = \frac{\rho_{\rm DE}(z_{\rm ta})}{\zeta \rho_m(z_{\rm ta})} \ .
\end{eqnarray}
We have also introduced the ``deviation parameter'', $\eta$, as
\begin{eqnarray}
\label{q}
\eta = \frac{\langle \, [1+3w(z)] \, \rho_{\rm DE}(z) \, \rangle}{(1+3w_{\rm ta}) \, \rho_{\rm DE}(z_{\rm ta})} \ .
\end{eqnarray}
It is worth noting that in the case of a cosmological constant
the system is conservative (see footnote 14), and the equation determining
$r_v/r_{\rm ta}$ is formally given by Eq.~(\ref{x}) with $\eta=1$. Hence,
the only restriction to the action of the $\langle ... \rangle$-operator is
that, when applied to the function $[1+3w(z)] \, \rho_{\rm DE}(z)$,
it must give $\eta=1$ for $w(z)=-1$.

Taking $\langle \, \psi(z) \, \rangle = \psi(z_{\rm ta})$
corresponds to the choice of Wang, which also implies
$\eta=1$ for equation-of-state parameter $w(z)$.
However, taking $\langle \, \psi(z) \, \rangle = \psi(\bar{z})$,
where $\bar{z}$ can be anywhere in the interval $[z_v,z_{\rm ta}]$,
is also a plausible choice.


\begin{figure}[t]
\begin{center}
\vspace*{-1.08cm}
\hspace*{-1.9cm}
\includegraphics[clip,width=0.66\textwidth]{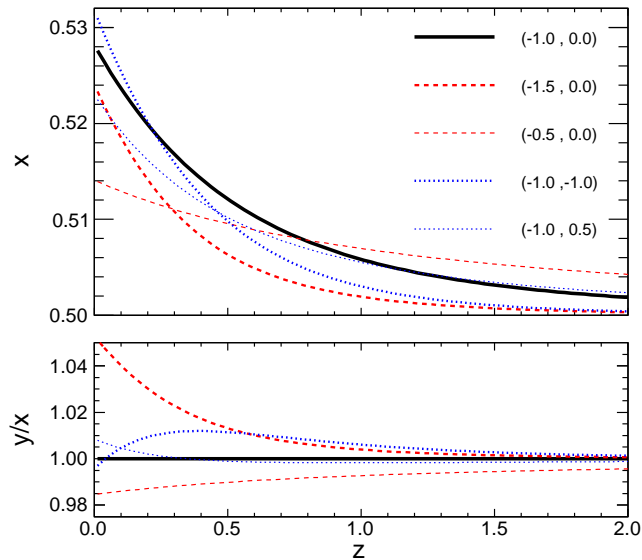}
\vspace*{-2.8cm}
\caption{{\it Upper panel}. The ratio of the virial radius to the turn-around
radius, $x \equiv r_v/r_{\rm ta}$
with $\eta = 1$ (corresponding to the Wang's choice),
as a function of the redshift for different values of $(w_0,w_a)$
(the same as in Fig.~1).
{\it Lower panel}. The ratio of $x$ to $y$, where $y \equiv r_v/r_{\rm ta}$ with
$\eta$ defined in Eq.~(\ref{xy}).
In both panels, we fixed $\Omega_m = 0.27$.}
\end{center}
\end{figure}


%

In the upper panel of Fig.~9, we plot the ratio of the virial radius to the turn-around
radius in the case $\eta=1$ (Wang's choice) as a function of the redshift for different
values of $(w_0,w_a)$.
In the lower panel, we show the same ratio for the case $\langle \, \psi(z) \, \rangle = \psi(z_v)$.
For notational clarity, we indicate those ratios by
\begin{equation}
\label{xy}
\frac{r_v}{r_{\rm ta}} \equiv
\left\{ \begin{array}{ll}
   x & \; \mbox{if} \;\; \eta = 1 \ ,
\\
\\
   y & \; \mbox{if} \;\; \eta = \frac{(1+3w_v) \, \rho_{\rm DE}(z_v)}{(1+3w_{\rm ta}) \, \rho_{\rm DE}(z_{\rm ta})} \ ,
\end{array}
\right.
\end{equation}
where $w_v = w(z_v)$.
For $z \gg 1$, where dark energy effects can be neglected,
the ratio of the virial to the turn-around radii approaches,
as expected, the asymptotic Einstein-de Sitter value $x=1/2$.
The variations of $x$ with the choice of the functional form
of the deviation parameter $\eta$ are of order of a few percent.
The resulting analysis on the growth of massive galaxy clusters
does not appreciably depend on $\eta$, and
the results presented in the previous sections are for the case $\eta=1$.



\begin{thebibliography}{99}


\bibitem{Review}            S.\ W.\ Allen, A.\ E.\ Evrard and A.\ B.\ Mantz,
                            arXiv\-:1103.4829 [astro-ph.CO].

\bibitem{Peebles84}         P.\ J.\ E.\ Peebles,
                            Astrophys.\ J. {\bf 284}, 439 (1984).

\bibitem{Weinberg}          S.\ Weinberg,
                            {\it Cosmology}
                            (Oxford University Press, New York, New York, 2008).

\bibitem{Voit}              G.\ M.\ Voit,
                            Rev.\ Mod.\ Phys.\ {\bf 77}, 207 (2005)
                            [arXiv:astro-ph/0410173].

\bibitem{Dark}              P.\ Brax, arXiv:0912.3610 [astro-ph.CO];
                            J.-P.\ Uzan, ar\-Xiv:0912\-.5452 [gr-qc];
                            A.\ De Felice and S.\ Tsujikawa, Living Rev.\ Rel.\ {\bf 13}, 3 (2010)
                            [arXiv:1002.4928 [gr-qc]];
                            E.\ V.\ Linder, arXiv:1004.4646 [astro-ph.CO];
                            A.\ Blanchard, Astron.\ Astrophys.\ Rev.\ {\bf 18}, 595 (2010)
                            [arXiv:1005.3765 [astro-ph.CO]];
                            D.\ Sapone, Int.\ J.\ Mod.\ Phys.\ A {\bf 25}, 5253 (2010)
                            [arXiv:1006.5694 [astro-ph.CO]].

\bibitem{LambdaObs}         H.\ K.\ Jassal, J.\ S.\ Bagla and T.\ Padmanabhan,
                            Mon.\ Not.\ Roy.\ Astron.\ Soc.\ {\bf 405}, 2639 (2010)
                            [arXiv:astro-ph/0601389];
                            K.\ M.\ Wilson, G.\ Chen and B.\ Ratra,
                            Mod.\ Phys.\ Lett.\ A {\bf 21}, 2197 (2006)
                            [arXiv:astro-ph/0602321];
                            T.\ M.\ Davis {\it et al.},
                            Astrophys.\ J.\ {\bf 666}, 716 (2007)
                            [arXiv:astro-ph/0701510].

\bibitem{CDMtrouble}        P.\ J.\ E.\ Peebles and B.\ Ratra,
                            Rev.\ Mod.\ Phys.\ {\bf 75}, 559 (2003)
                            [arXiv:astro-ph/0207347];
                            L.\ Perivolaropoulos,
                            J.\ Phys.\ Conf.\ Ser.\ {\bf 222}, 012024 (2010)
                            [arXiv:1002.3030 [astro-ph.CO]].

\bibitem{PR88}              P.\ J.\ E.\ Peebles and B.\ Ratra,
                            Astrophys.\ J.\ Lett.\ {\bf 325}, L17 (1988).

\bibitem{RP88}              B.\ Ratra and P.\ J.\ E.\ Peebles,
                            Phys.\ Rev.\ D {\bf 37}, 3406 (1988).

\bibitem{Mantz1}            A.\ Mantz, S.\ W.\ Allen, H.\ Ebeling and D.\ Rapetti,
                            Mon.\ Not.\ Roy.\ Astron.\ Soc.\ {\bf 387}, 1179 (2008)
                            [arXiv:0709.4294 [astro-ph]].

\bibitem{Mantz2}            A.\ Mantz, S.\ W.\ Allen, D.\ Rapetti and H.\ Ebeling,
                            Mon.\ Not.\ Roy.\ Astron.\ Soc.\ {\bf 406}, 1759 (2010)
                            [arXiv:0909.3098 [astro-ph.CO]].

\bibitem{Vikhlinin}         A.\ Vikhlinin {\it et al.},
                            Astrophys.\ J.\ {\bf 692}, 1060 (2009)
                            [arXiv:0812.2720 [astro-ph]].

\bibitem{K12}                N.\ A.\ Arhipova, T.\ Kahniashvili and V.\ N.\ Lukash,
                            Astr\-on.\ Astrophys.\ {\bf 386}, 775 (2002)
                            [arXiv:astro-ph/\-0110426];
                            T.\ Kahniashvili, E.\ von Toerne, N.\ A.\ Arhi\-pova and B.\ Ratra,
                            Phys.\ Rev.\ D {\bf 71}, 125009 (2005)
                            [arXiv:astro-ph/0503328].

\bibitem{Rozo}              E.\ Rozo {\it et al.},
                            Astrophys.\ J.\ {\bf 708}, 645 (2010)
                            [arXiv\-:0902.3702 [astro-ph.CO]].

\bibitem{Basilakos}         S.~Basilakos, M.~Plionis and J.~A.~S.~Lima,
                            Phys.\ Rev.\ D {\bf 82}, 083517 (2010)
                            [arXiv:1006.3418 [astro-ph.CO]];
                            M.~Manera and D.~F.~Mota,
                            Mon.\ Not.\ Roy.\ Astron.\ Soc.\ {\bf 371}, 1373 (2006)
                            [astro-ph/0504519];
                            S.~Basilakos, M.~Plionis and J.~Sola,
                            Phys.\ Rev.\ D {\bf 80}, 083511 (2009)
                            [arXiv:0907.4555 [astro-ph.CO]];
                            J.~Grande, J.~Sola, S.~Basilakos and M.~Plionis,
                            JCAP {\bf 1108}, 007 (2011)
                            [arXiv:1103.4632 [astro-ph.CO]].

\bibitem{Bahcall}           N.\ A.\ Bahcall and X.-h.\ Fan,
                            Astrophys.\ J.\ {\bf 504}, 1 (1998)
                            [arXiv:astro-ph/9803277];
                            N.\ A.\ Bahcall and P.\ Bode,
                            Astrophys.\ J.\ {\bf 588}, L1 (2003)
                            [arXiv:astro-ph/0212363].

\bibitem{CP}                M.\ Chevallier and D.\ Polarski,
                            Int.\ J.\ Mod.\ Phys.\ D {\bf 10}, 213 (2001)
                            [arXiv:gr-qc/0009008];
                            E.\ V.\ Linder,
                            Phys.\ Rev.\ Lett.\ {\bf 90}, 091301 (2003)
                            [arXiv:astro-ph/0208512].
                                                
\bibitem{Abramo}            L.~R.~Abramo, R.~C.~Batista and R.~Rosenfeld,
                            JCAP {\bf 0907}, 040 (2009)
                            [arXiv:0902.3226 [astro-ph.CO]].

\bibitem{Podariu01}         S.\ Podariu, T.\ Souradeep, J.\ R.\ Gott, B.\ Ratra and M.\ S.\ Vogeley,
                            Astrophys.\ J.\ {\bf 559}, 9 (2001)
                            [arXiv:astro-ph/0102264].

\bibitem{WMAP7}             E.\ Komatsu {\it et al.}  [WMAP Collaboration],
                            Astrophys.\ J.\ Suppl.\ {\bf 192}, 18 (2011)
                            [arXiv:1001.4538 [astro-ph.CO]].

\bibitem{PS}                W.\ H.\ Press and P.\ Schechter,
                            Astrophys.\ J.\ {\bf 187}, 425 (1974).

\bibitem{ST}                R.\ K.\ Sheth and G.\ Tormen,
                            Mon.\ Not.\ Roy.\ Astron.\ Soc.\ {\bf 308}, 119 (1999)
                            [arXiv:astro-ph/9901122].

\bibitem{Hu}                D.\ J.\ Eisenstein and W.\ Hu,
                            Astrophys.\ J.\ {\bf 496}, 605 (1998)
                            [arXiv:astro-ph/9709112].

\bibitem{Ikebe}             Y.\ Ikebe, T.\ H.\ Reiprich, H.\ Boehringer, Y.\ Tanaka and T.\ Kitayama,
                            Astron.\ Astrophys.\ {\bf 383}, 773 (2002)
                            [arXiv:astro-ph/0112315].

\bibitem{Henry}             J.\ P.\ Henry,
                            Astrophys.\ J.\ {\bf 534}, 565 (2000)
                            [arXiv:astro-ph/0002365].

\bibitem{Schmidt}           M.\ Schmidt,
                            Astrophys.\ J.\ {\bf 151}, 393 (1968);
                            Y.\ Avni and J.\ N.\ Bahcall,
                            Astrophys.\ J.\ {\bf 235}, 694 (1980).

\bibitem{Donahue2}          M.\ Donahue, G.\ M.\ Voit, I.\ M.\ Gioia, G.\ Luppino, J.\ P.\ Hughes and
                            J.\ T.\ Stocke,
                            arXiv:astro-ph/9707010. 

\bibitem{Hjorth}            J.\ Hjorth, J.\ Oukbir and E.\ van Kampen,
                            New Astron.\ Rev.\ {\bf 42}, 145 (1998);
                            Mon.\ Not.\ Roy.\ Astron.\ Soc.\ {\bf 298}, L1 (1998)
                            [arXiv:astro-ph/9802293].

\bibitem{T1}                P.\ E.\ J.\ Nulsen, S.\ L.\ Powell and A.\ Vikhlinin,
                            Astrophys.\ J.\ {\bf 722}, 55 (2010)
                            [arXiv:1008.2393 [astro-ph.CO]].

\bibitem{Getting}           G.\ L.\ Fogli, E.\ Lisi, A.\ Marrone, D.\ Montanino and A.\ Palazzo,
                            Phys.\ Rev.\ D {\bf 66}, 053010 (2002)
                            [arXiv:hep-ph/0206162].

\bibitem{Gaztanagaetal}     E.\ Gazta\~naga, A.\ Cabr\'e and L.\ Hui,
                            Mon.\ Not.\ Roy.\ Astron.\ Soc.\ {\bf 399}, 1663 (2009)
                            [arXiv:0807.3551 [astro-ph]];
                            L.\ Samushia and B.\ Ratra,
                            Astrophys.\ J.\ {\bf 701}, 1373 (2009)
                            [arXiv:0810.2104 [astro-ph]];
                            Y. Wang, Mod.\ Phys.\ Lett.\ A {\bf 25}, 3093 (2010)
                            [arXiv:0904.2218 [astro-ph.CO]],
                            and references therein.

\bibitem{Percival}          W.\ J.\ Percival {\it et al.}  [SDSS Collaboration],
                            Mon.\ Not.\ Roy.\ Astron.\ Soc.\ {\bf 401}, 2148 (2010)
                            [arXiv:0907.1660 [astro-ph.CO]].

\bibitem{Kolb}              E.\ W.\ Kolb and M.\ S.\ Turner,
                            {\it The Early Universe}
                            (Addison-Wesley, Redwood City, California, 1990).

\bibitem{Melchiorri}        P.~S.~Corasaniti and A.~Melchiorri,
                            Phys.\ Rev.\ D {\bf 77}, 103507 (2008)
                            [arXiv:0711.4119 [astro-ph]].

\bibitem{Ratra}             G.\ Chen and B.\ Ratra,
                            Publ.\ Astron.\ Soc.\ Pacific {\bf 123}, 1127 (2011)
                            [arXiv:1105.5206 [astro-ph.CO]]; also see
                            G.\ Chen, J.\ R.\ Gott and B.\ Ratra,
                            Publ.\ Astron.\ Soc.\ Pacific {\bf 115}, 1269 (2003)
                            [arXiv:astro-ph/0308099].

\bibitem{Tammann}           G.\ A.\ Tamman, A.\ Sandage and B.\ Reindl,
                            Astron.\ Astrophys.\ Rev.\ {\bf 15}, 289 (2008)
                            [arXiv:0806.3018 [astro-ph]].

\bibitem{Freedman}          W.\ L.\ Freedman and B.\ F.\ Madore,
                            Ann.\ Rev.\ Astron.\ Astrophys.\ {\bf 48}, 673 (2010)
                            [arXiv:1004.1856 [astro-ph.CO]].

\bibitem{Jimenez}           R.\ Jimenez and A.\ Loeb,
                            Astrophys.\ J.\ {\bf 573}, 37 (2002)
                            [arXiv:astro-ph/0106145].

\bibitem{Stern}             D.\ Stern, R.\ Jimenez, L.\ Verde, M.\ Kamionkowski and S.\ A.\ Stanford,
                            JCAP {\bf 1002}, 008 (2010)
                            [arXiv:0907.3149 [astro-ph.CO]].

\bibitem{Gaztanaga}         E.\ Gazta\~naga, A.\ Cabr\'e and L.\ Hui,
                            Mon.\ Not.\ Roy.\ Astron.\ Soc.\ {\bf 399}, 1663 (2009)
                            [arXiv:0807.3551 [astro-ph]].

\bibitem{hubbleparameter}   L.\ Samushia and B.\ Ratra,
                            Astrophys.\ J.\ Lett.\ {\bf 650}, L5 (2006)
                            [arXiv:astro-ph/0607301];
                            H.\ Zhang and Z.-H.\ Zhu,
                            JCAP {\bf 0803}, 007 (2008)
                            [arXiv:astro-ph/0703245];
                            A.\ A.\ Sen and R.\ J.\ Scherrer,
                            Phys.\ Lett.\ B {\bf 659}, 457 (2008)
                            [arXiv:astro-ph/0703416];
                            L.\ Samushia, G.\ Chen and B.\ Ratra,
                            arXiv:0706.1963 [astro-ph];
                            N.\ Pan, Y.\ Gong, Y.\ Chen and Z.-H.\ Zhu,
                            Class.\ Quantum Grav.\ {\bf 27}, 155015 (2010)
                            [arXiv:1005.4249 [astro-ph.CO]];
                            Y.\ Chen and B.\ Ratra,
                            Phys.\ Lett.\ B {\bf 703}, 406 (2011)
                            [arXiv:1106.4294 [astro-ph.CO]], and references therein.

\bibitem{Union2}            R.\ Amanullah {\it et al.},
                            Astrophys.\ J.\ {\bf 716}, 712 (2010)
                            [arXiv:1004.1711 [astro-ph.CO]].

\bibitem{SALT2}             J.\ Guy {\it et al.},
                            Astron.\ Astrophys.\ {\bf 466}, 11 (2007)
                            [arXiv\-:astro-ph/0701828].

\bibitem{Lewis}             A.\ Lewis and S.\ Bridle,
                            Phys.\ Rev.\ D {\bf 66}, 103511 (2002)
                            [arXiv:astro-ph/0205436].

\bibitem{Marrone}           L.\ Campanelli, P.\ Cea, G.\ L.\ Fogli and A.\ Marrone,
                            Phys.\ Rev.\ D {\bf 83}, 103503 (2011)
                            [arXiv:1012.5596 [astro-ph.CO]].

\bibitem{GL}                K.-H.\ Chae, G.\ Chen, B.\ Ratra and D.-W.\ Lee,
                            Astrophys.\ J.\ Lett.\ {\bf 607}, L71 (2004)
                            [arXiv:astro-ph/0403256];
                            S.\ Lee and K.-W.\ Ng,
                            Phys.\ Rev.\ D {\bf 76}, 043518 (2007)
                            [arXiv:0707.1730 [astro-ph]];
                            M.\ Yashar, B.\ Bozek, A.\ Abrahamse, A.\ Albrecht and M.\ Barnard,
                            Phys.\ Rev.\  D {\bf 79}, 103004 (2009)
                            [arXiv:0811.2253 [astro-ph]];
                            M.\ Biesiada, A.\ Piorkowska and B.\ Malec,
                            Mon.\ Not.\ Roy.\ Astron.\ Soc.\  {\bf 406}, 1055 (2010)
                            [arXiv:1105.0946 [astro-ph.CO]], and references therein.

\bibitem{PDB}               S.~Eidelman {\it et al.}  [Particle Data Group],
                            Phys.\ Lett.\ B {\bf 592}, 1 (2004).

\bibitem{Schuecker}         P.\ Schuecker, H.\ Bohringer, C.\ A.\ Collins and L.\ Guzzo,
                            Astron.\ Astrophys.\ {\bf 398}, 867 (2003)
                            [arXiv:astro-ph/\-0208251].

\bibitem{Riess}             A.\ G.\ Riess {\it et al.},
                            Astrophys.\ J.\ {\bf 730}, 119 (2011)
                            [Erratum-ibid.\ {\bf 732}, 129 (2011)]
                            [arXiv:1103.2976 [astro-ph.CO]].

\bibitem{ADD}               L.\ I.\ Gurvits, K.\ I.\ Kellermann, and S.\ Frey,
                            Astron.\ Astrophys.\ {\bf 342}, 378 (1999)
                            [arXiv:astro-ph/9812018];
                            E.\ J.\ Guerra, R.\ A.\ Daly, and L.\ Wan,
                            Astrophys.\ J.\ {\bf 544}, 659 (2000)
                            [arXiv:astro-ph/0006454];
                            G.\ Chen and B.\ Ratra,
                            Astrophys.\ J.\  {\bf 582}, 586 (2003)
                            [arXiv:astro-ph/0207051];
                            M.\ Bonamente, M.\ K.\ Joy, S.\ J.\ La Roque, J.\ E.\ Carlstrom,
                            E.\ D.\ Reese and K.\ S.\ Dawson,
                            Astrophys.\ J.\  {\bf 647}, 25 (2006)
                            [arXiv:astro-ph/0512349];
                            Y.\ Chen and B.\ Ratra,
                            arXiv:1105.5660 [astro-ph.CO], and references therein.

\bibitem{LT}                S.\ Capozziello, V.\ F.\ Cardone, M.\ Funaro and S.\ Andreon,
                            Phys.\ Rev.\  D {\bf 70}, 123501 (2004)
                            [arXiv:astro-ph/0410268];
                            N.\ Pires, Z.-H.\ Zhu and J.\ S.\ Alcaniz,
                            Phys.\ Rev.\  D {\bf 73}, 123530 (2006)
                            [arXiv:astro-ph/0606689];
                            L.\ Samushia, A.\ Dev, D.\ Jain and B.\ Ratra,
                            Phys.\ Lett.\  B {\bf 693}, 509 (2010)
                            [arXiv:0906.2734 [astro-ph.CO]];
                            M.\ A.\ Dantas, J.\ S.\ Alcaniz, D.\ Mania and B.\ Ratra,
                            Phys.\ Lett.\  B {\bf 699}, 239 (2011)
                            [arXiv:1010.0995 [astro-ph.CO]], and references therein.

\bibitem{GMF}               S.\ W.\ Allen, D.\ A.\ Rapetti, R.\ W.\ Schmidt, H.\ Ebeling, G.\ Morris
                            and A.\ C.\ Fabian,
                            Mon.\ Not.\ Roy.\ Astron.\ Soc.\  {\bf 383}, 879 (2008)
                            [arXiv:0706.0033 [astro-ph]];
                            L.\ Samushia and B.\ Ratra,
                            Astrophys.\ J.\  Lett.\ {\bf 680}, L1 (2008)
                            [arXiv:0803.3775 [astro-ph]];
                            S.\ Ettori, {\it et al.},
                            Astron.\ Astrophys.\  {\bf 501}, 61 (2009)
                            [arXiv:0904.2740 [astro-ph.CO]], and references therein.

\bibitem{GRB}               B.\ E.\ Schaefer,
                            Astrophys.\ J.\  {\bf 660}, 16 (2007)
                            [arXiv:astro-ph/0612285];
                            N.\ Liang and S.\ N.\ Zhang,
                            AIP Conf.\ Proc.\  {\bf 1065}, 367 (2008)
                            [arXiv:0808.2655 [astro-ph]];
                            Y.\ Wang,
                            Phys.\ Rev.\  D {\bf 78}, 123532 (2008)
                            [arXiv:0809.0657 [astro-ph]];
                            L.\ Samushia and B.\ Ratra,
                            Astrophys.\ J.\  {\bf 714}, 1347 (2010)
                            [arXiv:0905.3836 [astro-ph.CO]], and references therein.

\bibitem{Barger}            V.\ Barger, E.\ Guarnaccia and D.\ Marfatia,
                            Phys.\ Lett.\ B {\bf 635}, 61 (2006)
                            [arXiv:hep-ph/0512320].

\bibitem{Cadwell}           R.\ R.\ Caldwell and E.\ V.\ Linder,
                            Phys.\ Rev.\ Lett.\ {\bf 95}, 141301 (2005)
                            [arXiv:astro-ph/0505494].

\bibitem{Gupta}             G.~Gupta, S.~Majumdar and A.~A.~Sen,
                            Mon.\ Not.\ Roy.\ Astron.\ Soc.\ {\bf 420}, 1309 (2012)
                            [arXiv:1109.4112 [astro-ph.CO]].

\bibitem{Sherrer}           R.\ J.\ Scherrer,
                            Phys.\ Rev.\ D {\bf 73}, 043502 (2006)
                            [arXiv\-:astro-ph/0509890].

\bibitem{Mukhanov}          C.\ Armendariz-Picon, V.\ F.\ Mukhanov and P.\ J.\ Steinhardt,
                            Phys.\ Rev.\ Lett.\ {\bf 85}, 4438 (2000)
                            [arXiv:astro-ph/0004134].

\bibitem{Chiba}             T.\ Chiba,
                            Phys.\ Rev.\ D {\bf 73}, 063501 (2006)
                            [Erratum-ibid.\ D {\bf 80}, 129901 (2009)]
                            [arXiv:astro-ph/0510598].

\bibitem{Moschella}         A.\ Y.\ Kamenshchik, U.\ Moschella and V.\ Pasquier,
                            Phys.\ Lett.\ B {\bf 511}, 265 (2001)
                            [arXiv:gr-qc/0103004].

\bibitem{Bertolami}         M.\ C.\ Bento, O.\ Bertolami and A.\ A.\ Sen,
                            Phys.\ Rev.\ D {\bf 66}, 043507 (2002)
                            [arXiv:gr-qc/0202064].

\bibitem{XCDMincompleteness} B.\ Ratra,
                             Phys.\ Rev.\ D {\bf 43}, 3802 (1991);
                             S.\ Podariu and B.\ Ratra,
                             Astrophys.\ J.\ {\bf 532}, 109 (2000)
                             [arXiv:astro-ph/991057].

\bibitem{Pace}              F.\ Pace, J.\ C.\ Waizmann and M.\ Bartelmann,
                            arXiv\-:1005.0233 [astro-ph.CO].

\bibitem{Mota}              D.~F.~Mota,
                            JCAP {\bf 0809}, 006 (2008)
                            [arXiv:0812.4493 [astro-ph]].

\bibitem{NFW}               J.\ F.\ Navarro, C.\ S.\ Frenk and S.\ D.\ M.\ White,
                            Mon.\ Not.\ Roy.\ Astron.\ Soc.\ {\bf 275}, 720 (1995)
                            [arXiv:astro-ph/9408069].

\bibitem{Bullock}           J.\ S.\ Bullock {\it et al.},
                            Mon.\ Not.\ Roy.\ Astron.\ Soc.\ {\bf 321}, 559 (2001)
                            [arXiv:astro-ph/9908159].

\bibitem{Wang}              P.\ Wang,
                            Astrophys.\ J.\ {\bf 640}, 18 (2006)
                            [arXiv:astro-ph/0507195].

\end{thebibliography}
\end{document}